\documentclass[superscriptaddress,aps,preprintnumbers,amsmath,amssymb,prd,nofootinbib,preprint,longbibliography]{revtex4-1}
\pdfoutput=1

\usepackage{epsfig}
\usepackage{amsmath}
\usepackage{amssymb}
\usepackage{slashed}
\usepackage{verbatim}
\usepackage{bm}
\usepackage{dsfont}
\usepackage{cancel}
\usepackage{hyperref}
\usepackage{xcolor}
\usepackage{hyperref}
\usepackage[normalem]{ulem}
\usepackage{setspace}
\hypersetup{
colorlinks = true,
linkcolor=blue,
citecolor=green!40!black,
urlcolor=green!40!black,
}

\usepackage{multirow}
\usepackage{bbold}

\numberwithin{equation}{section}

\usepackage{soul}

\newcommand{\MeV}{\textrm{ MeV}}
\newcommand{\GeV}{\textrm{ GeV}}
\newcommand{\TeV}{\textrm{ TeV}}
\newcommand{\PeV}{\textrm{ PeV}}
\newcommand{\sx}{\sigma}

\begin{document}
\count\footins = 1000 

\begin{flushright}
UMN--TH--4206/22, FTPI--MINN--22/36, CERN-TH-2022-199
\end{flushright}

\title{
Baryogenesis from Decaying Magnetic Helicity in Axiogenesis
}
\author{Raymond T.~Co}
\affiliation{William I. Fine Theoretical Physics Institute, School of Physics and Astronomy, University of Minnesota, Minneapolis, MN 55455, USA}
\author{Valerie Domcke}
\affiliation{Theoretical Physics Department, CERN, 1211 Geneva, Switzerland}
\author{Keisuke Harigaya}
\affiliation{Department of Physics, University of Chicago, Chicago, IL 60637, USA}
\affiliation{Enrico Fermi Institute, University of Chicago, Chicago, IL 60637, USA}
\affiliation{Kavli Institute for Cosmological Physics, University of Chicago, Chicago, IL 60637, USA}
\affiliation{Kavli Institute for the Physics and Mathematics of the Universe (WPI),
The University of Tokyo Institutes for Advanced Study,
The University of Tokyo, Kashiwa, Chiba 277-8583, Japan\vspace{1cm}}
\affiliation{Theoretical Physics Department, CERN, 1211 Geneva, Switzerland}

\vspace{0.5cm}

\begin{abstract}
Generating axion dark matter through the kinetic misalignment mechanism implies the generation of large asymmetries for Standard Model fermions in the early universe. Even if these asymmetries are washed out at later times, they can trigger a chiral plasma instability in the early universe.
Similarly, a direct coupling of the axion with the hypercharge gauge field can trigger a tachyonic instability.
These instabilities produce helical magnetic fields, which are preserved until the electroweak phase transition. At the electroweak phase transition, these become a source of baryon asymmetry, which can be much more efficient than the original axiogenesis proposal. We discuss constraints on axion dark matter production from the overproduction of the baryon asymmetry as well as a minimal, albeit fine-tuned setup, where both the correct dark matter abundance and baryon asymmetry can be achieved. For a given axion decay constant, this leads to a sharp prediction for the mass of the radial direction of the Peccei Quinn field, which is a soft mass scale in supersymmetric theories.
\end{abstract}

\maketitle

\tableofcontents

\newpage
\section{Introduction}

The QCD axion, introduced to address the strong $CP$ problem in QCD~\cite{Peccei:1977hh,Peccei:1977ur,Weinberg:1977ma,Wilczek:1977pj}, can simultaneously source axion dark matter and a baryon asymmetry through its dynamics in the early universe. A large initial velocity for the axion field, corresponding to an angular motion of the complex Peccei-Quinn (PQ) field or equivalently a non-zero PQ charge, can be initiated by the Affleck-Dine mechanism at large radial field values~\cite{Affleck:1984fy,Dine:1995kz}. The PQ charge is partly converted into a baryon asymmetry through strong and weak sphaleron processes~\cite{Co:2019wyp,Domcke:2020kcp,Co:2020xlh}. This is referred to as axiogenesis and is closely related to Affleck-Dine baryogenesis since the $U(1)$ charge of a complex field is transferred into baryon asymmetry, as well as to spontaneous baryogenesis~\cite{Cohen:1987vi,Cohen:1988kt} since the generation of the baryon asymmetry can be also understood in terms of the effective chemical potential provided by the axion velocity in combination with baryon-number violating processes in the Standard Model.

The kinetic energy of the axion field is converted into an axion abundance~\cite{Co:2019jts} (referred to as kinetic misalignment) which leads to an increased dark matter abundance in comparison with the standard misalignment mechanism~\cite{Preskill:1982cy,Abbott:1982af,Dine:1982ah}. This enhancement is due to the delay of the onset of the axion oscillations in the periodic potential which emerges around the QCD confinement scale. In addition, axion fluctuations can be produced through parametric resonance~\cite{Dolgov:1989us, Traschen:1990sw, Kofman:1994rk, Shtanov:1994ce, Kofman:1997yn}, referred to as axion fragmentation~\cite{Fonseca:2019ypl,Morgante:2021bks,Eroncel:2022vjg} (see also~\cite{Jaeckel:2016qjp, Berges:2019dgr}). Unfortunately, in this very minimal and predictive setup, the PQ charge necessary to obtain the observed baryon asymmetry leads to an overproduction of axion dark matter by a factor of at least $\mathcal{O}(100)$ unless either the weak anomaly coefficient of the PQ symmetry is large or the electroweak phase transition temperature is higher than predicted by the Standard Model~\cite{Co:2019wyp}.
See Refs.~\cite{Co:2020jtv,Harigaya:2021txz,Chakraborty:2021fkp,Kawamura:2021xpu,Co:2021qgl,Barnes:2022ren} for scenarios that solve this dark matter overproduction by additional interactions that enhance the baryon asymmetry produced by the axion velocity.

In this paper, we revisit the dynamics of the axion field in the early universe above the electroweak phase transition and note that the spontaneous $CP$-violation induced by the motion of the axion field leads to a chain of non-trivial phenomena.
The angular motion of the PQ field induces chemical potentials for the particles of the thermal bath. The resulting chiral chemical potential, describing the asymmetry between left- and right-handed particles, leads to an instability for one of the two helicity modes of the thermal hypermagnetic fields---a process referred to as the chiral plasma instability (CPI)~\cite{Joyce:1997uy} (see also \cite{Brandenburg:2017rcb,Schober:2017cdw,Schober:2018ojn}). Similarly, if present, the direct coupling of the axion with the hypercharge gauge field  induces a tachyonic instability for one of the two helicities of the gauge field~\cite{Turner:1987bw,Garretson:1992vt,Anber:2006xt}. 
If this helicity exceeds a critical value, turbulent processes in the plasma entail an inverse cascade during which the helicity is transferred to low-momentum modes and is approximately conserved until the electroweak phase transition~\cite{Brandenburg:2017rcb}. During the electroweak phase transition, the hypermagnetic helicity is converted into helical electromagnetic fields as well as some baryon asymmetry through the hypercharge anomaly of baryon symmetry~\cite{Giovannini:1997eg,Giovannini:1997gp,Kamada:2016eeb,Kamada:2016cnb}. Similar processes have been discussed for axion-like particles driving cosmic inflation~\cite{Anber:2009ua,Domcke:2018eki,Domcke:2019mnd,Domcke:2022kfs} and for rotating fields with a direct coupling to the hypercharge gauge field and heavy charged fermions due to a large initial Higgs field value~\cite{Kamada:2019uxp}.

We point out that this production of baryon asymmetry through the CPI is significantly more efficient than axiogenesis based on the weak anomaly of baryon symmetry for sufficiently large masses of the PQ field's radial direction. This has two important implications. Firstly, this excludes some of the parameter space of axion dark matter production through kinetic misalignment due to the overproduction of baryon asymmetry. In particular, this sets an upper bound on the charge density in the axion rotation, or equivalently an upper bound on the mass of the radial mode of the PQ field, which in supersymmetric theories translates to an upper bound on supersymmetry-breaking masses of scalar fields. Depending on the details of the axion model, we find this mass scale to be bounded by 10 TeV to 10 PeV. Similar constraints have recently been set on the chemical potentials associated with lepton-flavour asymmetries in the early universe in Ref.~\cite{Domcke:2022uue}.

Secondly, on the boundary of this exclusion, we can obtain the observed abundances of both dark matter and the baryon asymmetry, relying on the kinetic misalignment mechanism to generate axion dark matter around the QCD phase transition and on the CPI, along with decaying helical magnetic fields at the electroweak phase transition, to generate the baryon asymmetry. This scenario requires some fine-tuning to ensure that the CPI is only marginally efficient at the time when the radial mode of the PQ field settles to its minimum in order to generate the correct amount of baryon asymmetry. However, we stress the truly minimal nature of this proposal regarding the required field content: Within the familiar setup of, e.g., the DSFZ~\cite{Dine:1981rt,Zhitnitsky:1980tq} and KSVZ~\cite{Kim:1979if,Shifman:1979if} axion model, this solves the strong $CP$-problem and explains dark matter and the baryon asymmetry without the need of additional fields or new interactions to enhance the baryon-asymmetry production by the axion rotation. The fine-tuning can also be seen as a feature, giving a distinct and precise prediction for the mass of the radial mode of the PQ-field as a function of the axion decay constant.

The remainder of this paper is organized as follows. After reviewing the dynamics of the Peccei-Quinn field and the dark matter production through kinetic misalignment in Sec.~\ref{sec:KMM}, we describe the generation of the baryon asymmetry through the chiral plasma instability in Sec.~\ref{sec:BAU}. Our results are summarized in the figures presented in Sec.~\ref{sec:results}, followed by conclusions in Sec.~\ref{sec:conclusions}. Four appendices contain the technical details behind this work: App.~\ref{app:scaling} reviews the scaling relations in the kinetic misalignment mechanism, App.~\ref{app:c5} contains the computation of the coefficient relating the axion velocity and the chiral chemical potential in different temperature regimes and axion models, App.~\ref{app:helicity} illustrates how the minimization of the free energy determines the charge transfer from the axion velocity to the hypermagnetic helicity, and App.~\ref{app:reynolds} discusses the conditions for the onset of the inverse cascade, ensuring the survival of the helical magnetic fields until the electroweak phase transition.

\section{Axion dark matter production via kinetic misalignment}
\label{sec:KMM}

\subsection{Dynamics of the axion rotation}
\label{subsec:rotation}

We consider a PQ field $P$ with radial direction $r$ and angular (axion) direction $\theta$,
\begin{align}
\label{eq:P}
   P = \frac{1}{\sqrt{2}}r e^{i \theta/N_{\rm DW}},
\end{align}
in a wine-bottle type scalar potential with a minimum at $r = N_{\rm DW} f_a$, with $f_a$ the axion decay constant and $N_{\rm DW}$ the domain wall number.  Close to the minimum, the potential for the radial component is quadratic, $V(\sx) = m_\sx^2 \, \sx^2/2$, with $\sx = r - N_{\rm DW} f_a$ denoting the radial component degree of freedom in the true vacuum. At large field values $r \gg N_{\rm DW} f_a$, the scalar potential may be approximated by a quartic potential, $V(r) \simeq \lambda^2 r^4/4$ with $\lambda^2 = m_\sx^2/(2 N_{\rm DW}^2 f_a^2)$, as expected in a generic renormalizable theory, or may be approximately given by a quadratic potential, $V(r) \simeq m_\sx^2 \, r^2/2$, if the quartic component is suppressed. 
The latter is the case, e.g., in supersymmetric theories, which naturally give a flat potential for the radial mode  $\sx$ known as the saxion, lifted only by soft supersymmetry-breaking terms. In what follows, we will consider both possibilities.

We consider the situation where the axion is rapidly rotating at large field values, $\dot \theta \neq 0$ and $r \gg N_{\rm DW} f_a$. This can be achieved, for example, by the Affleck-Dine mechanism~\cite{Affleck:1984fy} but with higher-dimensional PQ-violating operators. Specifically, if $r$ is initially large, e.g., due to a negative Hubble induced mass during inflation~\cite{Dine:1995kz}, the higher dimensional PQ-violating operators then become effective and will provide a kick in the $\theta$ direction, resulting in an axion rotation.
The axion rotation carries a conserved PQ charge, and the charge yield is defined as the PQ charge density $n_\theta$ normalized to the entropy density $s = 2 \pi^2/45 \, g_*(T) T^3$, 
\begin{align}
\label{eq:Ytheta}
Y_\theta = \frac{n_\theta}{s} = \frac{\epsilon \, \omega \left( \frac{r_{\max}}{N_{\rm DW}}\right)^2 }{s} \hspace{1 cm} {\rm with} \hspace{1 cm} \omega^2 = N_{\rm DW}^2 \left. \frac{V'(r)}{r} \right|_{\max} .
\end{align}
The subscript `max' denotes the maximum value during a cycle. Here $0 < \epsilon < 1$ denotes the shape of the rotation in the complex plane with $\epsilon = 1$ ($\epsilon = 0$) corresponds to a perfectly circular motion (radial oscillation). In particular, the initial value of $\epsilon$ is given by the size of the initial kick (potential gradient) in the angular direction relative to the radial direction. The initial motion is necessarily elliptical $\epsilon < 1$ because the radial motion is also induced in the Affleck-Dine mechanism. However, upon thermalization of the PQ field with the thermal bath, the motion becomes perfectly circular ($\epsilon = 1$) because this is the state that minimizes the free energy of the thermal system for a fixed PQ charge, and the washout of the PQ charge by strong sphaleron processes and quark Yukawa interactions is slow~\cite{Co:2019wyp}.

After thermalization, $r$ is constant up to cosmic expansion, and therefore $\dot r \propto - H r$ with $H$ the Hubble expansion rate. Due to redshift, the Hubble scale becomes much smaller than the potential gradient in the radial direction soon after the onset of the motion. Using this approximation in the equation of motion of $r$, one finds that $\dot\theta^2 = N_{\rm DW}^2 V'(r)/r$, which can also be easily understood as the balance between the centripetal force $r (\dot\theta/N_{\rm DW} )^2$ and the the potential gradient $V'(r)$. From the equations of motion, one finds conservation of charge $Y_\theta = \text{constant}$ and can derive how various quantities scale.
When $r \gg N_{\rm DW} f_a$, one finds
\begin{align}
 r & \propto T \,, & |\dot \theta| & = \sqrt{2} N_{\rm DW} \lambda r \,, & \rho_\theta & \propto T^4 
 && \text{(quartic)} , \label{eq:BG>_4} \\
 r & \propto T^{3/2} \,, & |\dot \theta| & = N_{\rm DW} m_\sx \,, & \rho_\theta & \propto T^3  
 && \text{(quadratic)} ,
 \label{eq:BG>_2}
\end{align}
where $\rho_\theta = r^2 \left(\dot\theta/N_{\rm DW} \right)^2 / 2 + V(r)$ indicates the energy density of the rotation. Another way to derive these scaling laws is explained in App.~\ref{app:scaling}. 
For both the quartic and quadratic potentials, the radial field value decreases with cosmic expansion until settling to the true minimum $r = N_{\rm DW} f_a$ at the temperature denoted by $T_S$. After this point, one finds
\begin{align}
 \dot \theta \propto T^3 \,, \quad \rho_\theta \propto T^6 \; \,. 
 \label{eq:BG<}
\end{align}
The scaling of these quantities will be important to determine the onset of the CPI in Sec.~\ref{sec:BAU}.

\subsection{Axion dark matter from the kinetic misalignment mechanism}
\label{subsec:axion_DM}

This setup will give us axion dark matter (DM) production at $T < T_S$ around the QCD confinement scale through the kinetic misalignment mechanism~\cite{Co:2019jts}
\begin{align}
 \Omega_a h^2 = \Omega_\text{DM} h^2 \times c_\Omega \left( \frac{10^9~\text{GeV}}{f_a} \right) \left(\frac{Y_\theta}{73.3}\right) \, .
 \label{eq:DM}
\end{align}
Here $\Omega_\text{DM} h^2 = 0.12$~\cite{Planck:2018vyg} denotes the observed dark matter density and $c_\Omega$ is an order one fudge factor introduced to account for the conversion from the charge yield to the axion number yield. Such a conversion arises when the axion undergoes fragmentation~\cite{Fonseca:2019ypl}, where parametric resonance from axion self-interactions produces the axion fluctuations and destroys the coherent rotation. Analytic estimates give $c_\Omega \simeq 1$~\cite{ Co:2021rhi}, while numerical analyses result in $1 \lesssim  c_\Omega \lesssim 2$~\cite{Eroncel:2022vjg} for a regime that is not in the highly kinetic misalignment limit. For the present scenario, the axion is in the highly kinetic-misalignment regime and therefore the precise value of this factor $c_\Omega$ is unknown but expected to be of order unity.
Using $\left. Y_\theta \right|_{T_S} = |\dot\theta| f_a^2/(2\pi^2 g_* T_S^3 / 45)$ the requirement to reproduce the observed DM abundance $\Omega_\text{DM} h^2$ from kinetic misalignment at $T \ll T_S$ translates to
\begin{align}
\label{eq:TS}
 T_S & = \left( \frac{ N_{\rm DW} m_\sx f_a^2}{\frac{2\pi^2 g_*(T_S)}{45} Y_\theta} \right)^{\scalebox{1.01}{$\frac{1}{3}$}}
 \simeq 2.4 \PeV \, 
 c_\Omega^{1/3}
 \left( \frac{f_a}{10^9~\text{GeV}} \right)^{\scalebox{1.01}{$\frac{1}{3}$}}
 \left( \frac{N_{\rm DW}\,m_\sx}{10^5~\text{GeV}} \right)^{\scalebox{1.01}{$\frac{1}{3}$}}
 \left( \frac{g_{\rm MSSM}}{g_*(T_S)} \right)^{\scalebox{1.01}{$\frac{1}{3}$}} ,
\end{align}
where $g_{\rm MSSM} = 915/4$ denotes the number of degrees of freedom in the thermal bath at high temperatures in the minimal supersymmetric standard model (MSSM). This temperature $T_S$ separates the different scaling regimes as in Eqs.~\eqref{eq:BG>_4}--\eqref{eq:BG>_2}.

\subsection{Cosmological evolution}
\label{subsec:cosmo} 

In the case of a quadratic potential, Eqs.~\eqref{eq:BG>_2} and \eqref{eq:BG<} imply that the energy in the PQ sector $\rho_P$ can dominate over the thermal bath, leading to a matter-dominated era followed by a kination-dominated era. More precisely, the onset of this matter-dominated era occurs  when $\rho_{\rm th} = \rho_\theta = \dot\theta Y_\theta s$, or equivalently when $\pi^2 g_* T^4/30 = N_{\rm DW} m_\sx Y_\theta 2\pi^2 g_* T^3 / 45$. Using Eq.~(\ref{eq:DM}), we obtain the temperature at the transition of radiation to matter domination,
\begin{align}
    T_{\rm RM} = \frac{4 N_{\rm DW} m_\sx Y_\theta}{3} 
    \simeq 9.8 \PeV \;
    c_\Omega^{-1}
    \left( \frac{f_a}{10^9~\text{GeV}} \right)
    \left( \frac{N_{\rm DW}\,m_\sx}{10^5~\text{GeV}} \right) ,
\end{align}
if $T_{\rm RM} \gg T_S$.
After the matter-dominated era, a kination-dominated era commences at $T_S$. This kination-dominated era then transitions to a radiation-dominated one at temperature
\begin{align}
\label{eq:TKR}
    T_{\rm KR} 
    \simeq 1.7 \PeV \;
    c_\Omega
    \left( \frac{g_{\rm MSSM}}{g_*(T_{\rm KR})} \right)^{\scalebox{1.01}{$\frac{1}{2}$}} ,
\end{align}
obtained from $\rho_{\rm th} = \rho_\theta = n_\theta^2 / (2 f_a^2)$. Consistency in this case requires $T_S > T_{\rm KR}$, yielding
\begin{align}
 m_\sx \gtrsim  m_\sx^\text{kin} \equiv 34 \TeV \,  c_\Omega^2 N_{\rm DW}^{-1} 
 \left( \frac{10^9~\text{GeV}}{f_a} \right) 
 \left( \frac{g_{\rm MSSM}}{g_*(T_{\rm KR})} \right)^{\scalebox{1.01}{$\frac{3}{2}$}} 
 \left( \frac{g_*(T_S)}{g_{\rm MSSM}} \right) \,,
 \label{eq:kin}
\end{align}
as a condition for kination domination%
\footnote{Note that in deriving this condition, we would get a factor of 2 smaller $m_\sx^\text{kin}$ if we instead compare $T_{\rm RM}$ with $T_{\rm KR}$. This is because we assume that the rotation energy density consists of equal parts of the kinetic and potential contributions when deriving $T_{\rm RM}$, which is true for $r \gg N_\text{DW} f_a$. This is not true when $T_{\rm RM} \simeq T_S \simeq T_{\rm KR}$ since the potential energy approximately vanishes at $T_S$. \vspace{0.1cm}}%
, which will be shown as the magenta line in Fig.~\ref{fig:SUSY_TwoHiggs}.

For the quartic potential,
the energy density of the rotation cannot dominate over the radiation by redshifting according to Eq.~(\ref{eq:BG>_4}), but there is an upper bound on the charge yield from the requirement that $\rho_\theta$ should not dominate the total energy density.%
\footnote{The initial non-circular rotation can temporarily dominate the universe if the PQ field drives inflation, but thermalization entails a contribution to the radiation energy density which is at least comparable to the energy density in the circular rotation.}
Using $Y_\theta$ from Eq.~(\ref{eq:Ytheta}), $V(r) \simeq \lambda^2 r^4/4$, and $V(r) + \frac{1}{2} \epsilon r^2 V'(r)/r \le \pi^2 g_* T^4/30$, we can obtain the maximum charge yield and, together with Eq.~(\ref{eq:DM}), derive a constraint on the mass of $\sx$
\begin{align}
\label{eq:quartic_Ymax}
    m_\sx \lesssim 200 \TeV  \,
    c_\Omega^2 N_{\rm DW}^{-1}
    \frac{\epsilon^2}{\left(1+2\epsilon\right)^{3/2}} \left( \frac{10^9 \GeV}{f_a} \right)
    \left( \frac{g_{\rm SM}}{g_*} \right)^{\scalebox{1.01}{$\frac{1}{2}$}} ,
\end{align}
with $g_{\rm SM} = 427/4$ denoting the number of degrees of freedom in the Standard Model. 

In addition, for the quartic potential, there exists another upper bound on the charge yield from the requirement that $r$ should not already relax to small field values during inflation. Namely, we need to impose $\sqrt{V''(r)} < 3 H_I$ with $H_I$ the Hubble parameter during inflation so that the field is frozen at a large initial value due to Hubble friction despite an initial mass much larger than the vacuum one. To derive the maximum yield generated after inflation, we use the fact that the ratio $n_\theta/\rho_\varphi$ is redshift-invariant during a matter-dominated era lasting from the end of inflation to the end of reheating, where $\rho_\varphi$ is the inflaton energy density. With $n_\theta$ from Eq.~(\ref{eq:Ytheta}), we find the maximum charge yield after reheating
\begin{align}
    Y_\theta & = \left. \frac{n_\theta}{\rho_\varphi} \right|_{\rm end} \left. \frac{\rho_\varphi}{s} \right|_{T_R} \nonumber \\ 
    & \lesssim 1.8 \cdot 10^5 N_{\rm DW} \, \epsilon \,
    \left( \frac{H_I}{6 \times 10^{13} \GeV} \right)
    \left( \frac{T_R}{6.5 \times 10^{15} \GeV} \right)
    \left( \frac{f_a}{10^9 \GeV} \right)^2
    \left( \frac{\rm TeV}{m_\sx} \right)^2 , 
\end{align}
where, in the last inequality, the initial field value of $r$ is assumed to saturate the bound $\sqrt{V''(r)} < 3 H_I$. 
This maximum charge yield together with the DM requirement on $Y_\theta$ in Eq.~(\ref{eq:DM}) give an upper bound on $m_\sx$ as
\begin{align}
    m_\sx \lesssim 50 \TeV \ c_\Omega^{1/2} N_{\rm DW}^{1/2} \, \epsilon^{1/2} \,
    \left( \frac{H_I}{6 \times 10^{13} \GeV} \right)^{\scalebox{1.01}{$\frac{1}{2}$}}
    \left( \frac{T_R}{6.5 \times 10^{15} \GeV} \right)^{\scalebox{1.01}{$\frac{1}{2}$}}
    \left( \frac{f_a}{10^9 \GeV} \right)^{\scalebox{1.01}{$\frac{1}{2}$}}.
    \label{eq:quartic_Ymax_2}
\end{align}
This constraint is most relaxed for the largest $H_I$ and $T_R$. The benchmark values of $H_I$ and $T_R$ above are obtained by saturating the current bound on $H_I \lesssim 6 \times 10^{13} \GeV$ from {\it Planck}~2018~\cite{Planck:2018jri} and assuming instantaneous reheating after inflation. 
The exclusion by Eqs.~\eqref{eq:quartic_Ymax} and \eqref{eq:quartic_Ymax_2} will be shown as cyan regions in Figs.~\ref{fig:NonSUSY_TwoHiggs} and \ref{fig:OneHiggs}, where the former (latter) constraint corresponds to the cyan boundary with a negative (positive) slope.

\section{Baryogenesis due to chiral plasma instability}
\label{sec:BAU}

\subsection{Chiral plasma instability}
\label{subsec:CPI}

\begin{figure}[t]
\centering
 \includegraphics[width = 0.495 \textwidth]{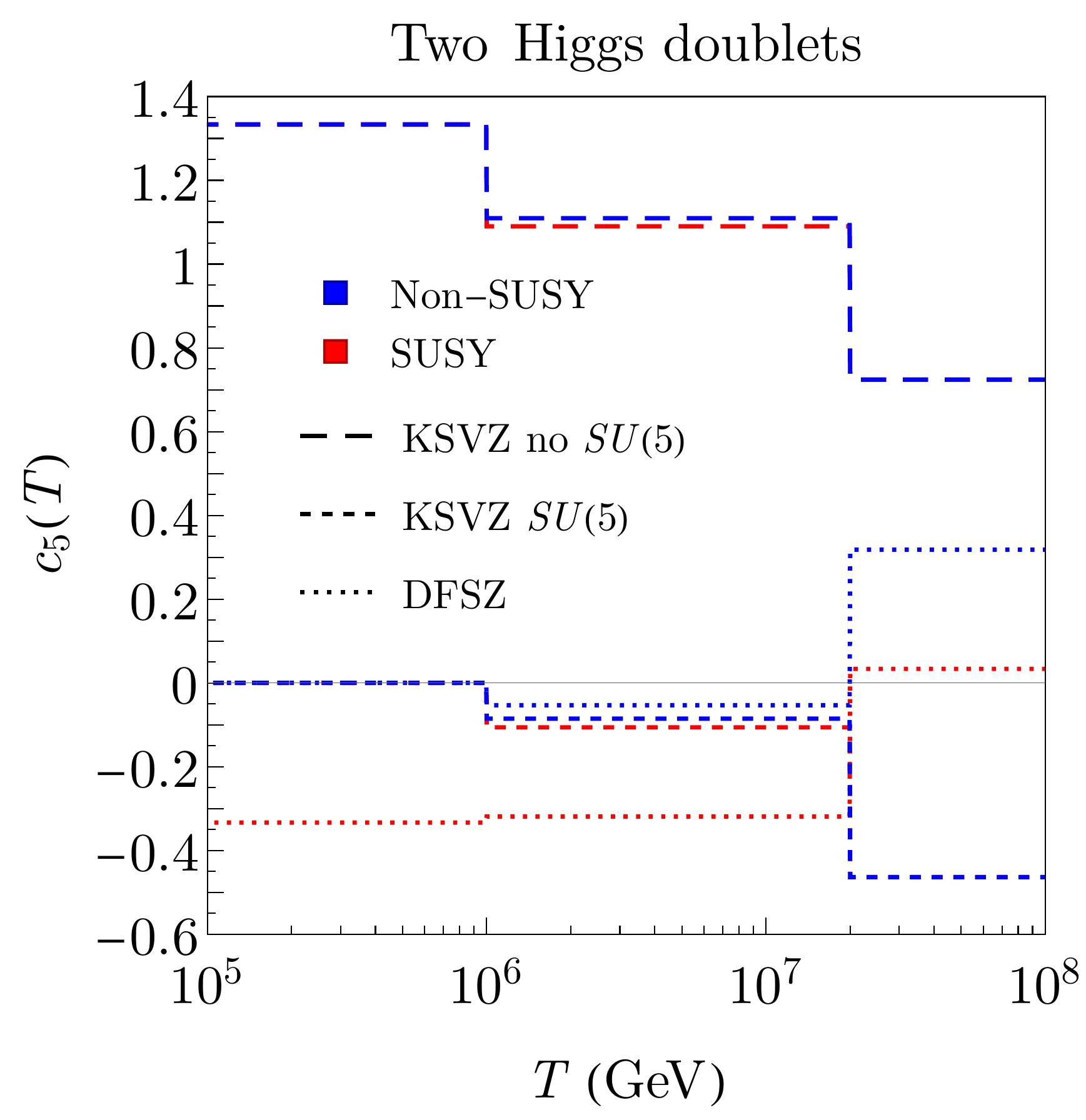} \hfill
 \includegraphics[width = 0.495 \textwidth]{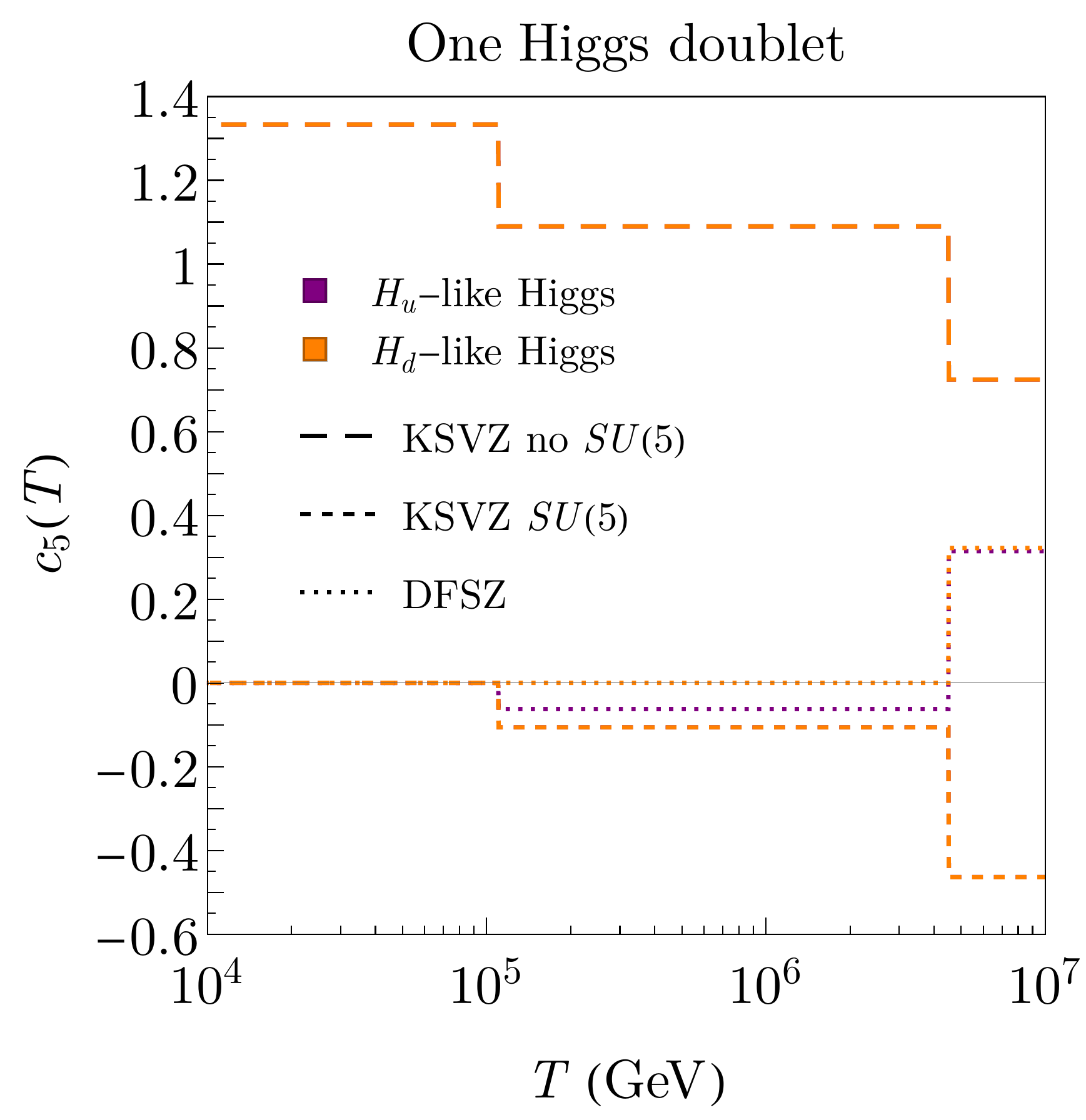}
 \caption{\footnotesize{The coefficient $c_5(T) \equiv \mu_{Y,5} / \dot\theta$ as functions of temperature. For models with two Higgs doublets shown in the left panel, we assume the electron (down) Yukawa comes into equilibrium instantaneously at $T = 10^6 \GeV$ ($T = 2 \times 10^7 \GeV$), corresponding to $\tan\beta \simeq 3$. For the one Higgs doublet models shown in the right panel, the electron (down) Yukawa comes into equilibrium at $T = 10^5 \GeV$ ($T = 4.5 \times 10^6 \GeV$). In both panels, the dotted lines are for the DFSZ axion model with $N_{\rm DW} = 3, n = 1, c_g=c_W=c_Y=0$, while the long-dashed (short-dashed) lines are for the KSVZ axion model, $N_{\rm DW} = 1, n = 0$, without (with) $SU(5)$ gauge unification so that $c_g=1, c_W=c_Y=0$ ($c_g=c_W=1, c_Y=5/3$). In the left panel, blue and red are for non-SUSY and SUSY, respectively. In the right panel, purple and orange are for the $H_u$-like and $H_d$-like Higgs, respectively, while the two cases are identical for the dashed lines. See App.~\ref{app:c5} for details.}} 
 \label{fig:c_5}
\end{figure}

The axion velocity together with the SM transport equations describing the Yukawa and sphaleron interactions generates chemical potentials for the SM fermions~\cite{Co:2019wyp,Domcke:2020kcp}. This will generically also generate a chiral chemical potential,
\begin{align}
 \mu_{Y,5} = \sum_i \varepsilon_i g_i Y_i^2 \mu_i = c_5 \, \dot \theta \,,
 \label{eq:mu5}
\end{align}
with $\varepsilon_i = \pm 1$ denoting right/left-handed particles, $g_i$ the multiplicity, and $Y_i$ the hypercharge of the (MS)SM particle species $i$. Consequently, $c_5$ denotes a temperature-dependent coefficient determined by the SM transport equations and the PQ model, which is typically ${\cal O}(0.01\mathchar`-1)$ and which we derive for various exemplary cases in App.~\ref{app:c5}; see also Fig.~\ref{fig:c_5}. There we also allow the contribution from the direct axion coupling with the hypercharge gauge field.

In particular, we consider supersymmetric (SUSY) and non-supersymmetric (non-SUSY) models in combination with both the KSVZ and the DSFZ axion models. As explained in more detail in App.~\ref{app:c5}, for supersymmetric models we assume all supersymmetric particles to be in the thermal bath and we will assume a quadratic potential for the radial mode of the PQ field. For non-supersymmetric models, we distinguish the type-II two Higgs-doublet extension of the SM as well as the SM with an $H_u$-like or $H_d$-like Higgs, and we will work with the quartic potential for the radial mode of the PQ field. For the DFSZ axion model at high temperatures when the Higgs fields are not integrated out, the axion does not couple to any of the gauge Chern-Simon terms but only to the Higgs fields. For the KSVZ axion model, the couplings to the Chern-Simons terms of the weak and hypercharge gauge fields depend on the gauge charge of the KSVZ fermions and we will work with two explicit examples---
the KSVZ fermions are and are not $SU(5)$-complete multiplets.

The chiral chemical potential~\eqref{eq:mu5} can trigger a chiral plasma instability~\cite{Joyce:1997uy}, which transfers the fermion asymmetries into helical hypermagnetic fields. 
This phenomenon is well studied in the context of chiral magnetohydrodynamics~\cite{Durrer:2013pga}. In the presence of a chiral chemical potential, the chiral magnetic effect~\cite{Fukushima:2008xe} leads to a contribution to the (hyper)electric current proportional to the (hyper)magnetic field. This gives rise to a tachyonic instability in one the two helicity modes of the gauge field, which results in a transfer of the fermion chiral asymmetry to helical gauge fields~\cite{Boyarsky:2011uy}. The underlying physics is in fact very similar to the tachyonic instability for helical gauge fields generated by a direct coupling to the axion~\cite{Turner:1987bw,Garretson:1992vt}. In this paper, we will use the term CPI to refer to both processes.
The onset of the CPI occurs when the typical time scale for the growth of fastest growing helicity mode%
\footnote{With $\mu_{Y,5}$ defined as in Eq.~\eqref{eq:mu5}, the contribution to hyperelectric current from the chiral magnetic effect reads ${\bf J}_Y \supset \alpha_Y \mu_{Y,5} {\bf B}_Y / \pi$. This is a factor 2 smaller than the expression given, e.g., in Refs.~\cite{Kamada:2018tcs,Domcke:2019mnd}.
We thank Kohei Kamada and Kyohei Mukaida for confirming this.}
$k_\text{CPI} /2= \alpha_Y \mu_{Y,5}/2\pi$ overcomes the Hubble expansion rate, i.e., $\Gamma_\text{CPI} > H$, with~\cite{Kamada:2018tcs}
\begin{align}
 \Gamma_\text{CPI} = \frac{k_\text{CPI}^2}{2 \sigma_Y} = \frac{\alpha_Y^2 \mu_{Y,5}^2 }{2\pi^2 \sigma_Y} \,,
 \label{eq:GammaCPI}
\end{align}
where $\alpha_Y \simeq 0.01$ denotes the hypercharge fine structure constant and $\sigma_Y$ denotes the conductivity before the electroweak phase transition (EWPT) with $\sigma_Y \simeq 54\,T$ for the SM, $\sigma_Y \simeq 50\,T$ for non-supersymmetric models with two Higgs doublets, and $\sigma_Y \simeq 40\,T$ for the MSSM~\cite{Arnold:2000dr}.%
\footnote{Here we only include the scattering of right-handed leptons and sleptons by the $U(1)_Y$ gauge interaction. In supersymmetric theories, slepton-lepton-bino interaction also contribute to the scattering so the actual conductivity can be smaller by an $\mathcal{O}(1)$ factor.}
We will use $\sigma_Y = 50T$ in all figures.
We moreover define a time scale for the CPI to become efficient, in particular to exponentially enhance the helical fields to a point where they will generate a baryon asymmetry comparable to the observed value (see below) through $\Gamma_\text{CPI} \equiv c_\text{CPI} H$ with $c_\text{CPI} > 1$. As we will see below, $c_\text{CPI} = {\cal O}(10)$.

For $T < T_S$, the CPI rate rapidly decreases as $\Gamma_\text{CPI} \propto T^5$, indicating that this process will be most efficient at $T \simeq T_S$.
We can estimate the CPI temperature $T_\text{CPI}$ at which $\Gamma_\text{CPI} = c_\text{CPI} H$ for $T > T_S$ as
\begin{align}
\label{eq:TCPI_quartic}
T_\text{CPI} & \simeq 8.5 \PeV 
\frac{c_5^2}{c_\Omega^{2/3}}
\left(\frac{10}{c_\text{CPI}} \frac{50\,T}{\sigma_Y}\right)
\left( \frac{ N_{\rm DW} \, m_\sx}{10^5~\text{GeV}} \right)^{\scalebox{1.01}{$\frac{4}{3}$}} 
\left( \frac{10^9~\text{GeV}}{f_a} \right)^{\scalebox{1.01}{$\frac{3}{2}$}} 
\left( \frac{g_*(T_{\rm CPI})}{g_{\rm MSSM}} \right)^{\scalebox{1.01}{$\frac{1}{6}$}} \,,
\end{align}
for the case of a quartic potential, and
\begin{align}
\label{eq:TCPI_quadratic_rad}
 T_\text{CPI}^\text{RD} & \simeq  3.6 \PeV
 |c_5|^{2/3} 
 \left(\frac{10}{c_\text{CPI}} \frac{50\,T}{\sigma_Y}\right)^{\scalebox{1.01}{$\frac{1}{3}$}}
 \left(\frac{N_{\rm DW}\,m_\sx}{10^5~\text{GeV}} \right)^{\scalebox{1.01}{$\frac{2}{3}$}} 
 \left( \frac{g_{\rm MSSM}}{g_*(T_{\rm CPI}^\text{RD})} \right)^{\scalebox{1.01}{$\frac{1}{6}$}} \,,\\ 
\label{eq:TCPI_quadratic_mat}
  T_\text{CPI}^\text{MD} & \simeq 3.0 \PeV \,  
  |c_5|^{4/5} \, c_\Omega^{1/5}
  \left(\frac{10}{c_\text{CPI}} \frac{50\,T}{\sigma_Y}\right)^{\scalebox{1.01}{$\frac{2}{5}$}}
  \left( \frac{N_{\rm DW}\,m_\sx}{10^5~\text{GeV}} \right)^{\scalebox{1.01}{$\frac{3}{5}$}} 
  \left( \frac{10^9~\text{GeV}}{f_a} \right)^{\scalebox{1.01}{$\frac{1}{5}$}} 
  \left( \frac{g_{\rm MSSM}}{g_*(T_{\rm CPI}^\text{MD})} \right)^{\scalebox{1.01}{$\frac{1}{5}$}} \,,
\end{align}
for the case of the quadratic potential, where the superscript RD (MD) indicates the assumption of radiation or matter domination for the quadratic potential at $T \sim T_\text{CPI} \geq T_S$.

The CPI never becomes relevant, i.e., $\Gamma_{\rm CPI} < c_\text{CPI} H$ at $T_S$, if 
\begin{align}
 m_\sx \lesssim 28 \TeV \; 
 \frac{c_\Omega}{c_5^2 \, N_{\rm DW}}
 \left( \frac{c_\text{CPI}}{10} \frac{\sigma_Y}{50\,T} \right) 
 \left(\frac{f_a}{10^9~\text{GeV}}\right)  
 \left( \frac{g_{\rm MSSM}}{g_*(T_S)} \right)^{\scalebox{1.01}{$\frac{1}{2}$}}
 \,, 
 \label{eq:mS_CPI}
\end{align}
 assuming radiation domination throughout, while the condition becomes
 \begin{align}
  m_\sx \lesssim 43 \TeV 
  \frac{c_\Omega^{1/2}}{|c_5|^{3} \, N_{\rm DW}}
  \left( \frac{c_\text{CPI}}{10} \frac{\sigma_Y}{50\,T}\right)^{\scalebox{1.01}{$\frac{3}{2}$}} 
  \left( \frac{f_a}{10^9~\text{GeV}} \right)^2  \,
  \left( \frac{g_{\rm MSSM}}{g_*(T_S)} \right)^{\scalebox{1.01}{$\frac{1}{2}$}} \,,
  \label{eq:mS_CPI_K}
\end{align}
if a phase of kination domination occurs.
Eqs.~\eqref{eq:mS_CPI} and \eqref{eq:mS_CPI_K} illustrate a key result of this paper. As we will see below, for values of $m_\sx$ above these bounds, the hypermagnetic helicity generated through the chiral plasma instability will entail an overproduction of the baryon asymmetry. The corresponding exclusion regions, taking into account the temperature dependence of $c_5(T)$, are shown as colored regions labeled $Y_B > Y_B^\text{obs}$ in Figs.~\ref{fig:SUSY_TwoHiggs}~to~\ref{fig:OneHiggs}.

If $\Gamma_{\rm CPI} > \mathcal{O}(10) H$ at $T_S$, the CPI becomes fully effective before the radial component of the PQ field reaches its minimum, i.e., $T_\text{CPI} > T_S$. (From Eqs.~\eqref{eq:mS_CPI}--\eqref{eq:mS_CPI_K} and Eq.~\eqref{eq:TS} or from the CPI regions along with the $T_S$ contours in Figs.~\ref{fig:SUSY_TwoHiggs} to \ref{fig:OneHiggs}, it is evident that $T_\text{CPI}, T_S \gg T_{\rm EW}$ in the entire parameter space relevant to the CPI. Namely, the magnetic helicity is generated well before EWPT.) Once the CPI is effective, the system will reach a quasi-equilibrium state that minimizes the free energy. As is argued in Appendix~\ref{app:helicity}, the helicity density $h$ at the quasi-equilibrium is
\begin{align}
  h \sim  \frac{\pi}{\alpha_Y} \dot{\theta} T^2 \, \text{sign}(c_5)\,.
    \label{eq:h}
\end{align}
The ${\cal O}(1)$ proportionality factor in this equation depends on temperature and is computed in App.~\ref{app:helicity} for the simplified case of only one fermion species. We note that  if all of the charge in the axion rotation were to be transferred into the helical magnetic field, $h$ would be ${ \cal O}(\dot{\theta} r^2 \pi/\alpha_Y)$.
The helicity density at the quasi-equilibrium is much smaller than this value by a factor $(T/r)^2$, which means that the axion rotation remains nearly intact.
This in particular implies that chiral chemical potential and hence the helicity can be sourced even when all SM interactions in the thermal plasma are efficient, i.e., even below the equilibrium temperature of the electron Yukawa coupling.
The sign of $\mu_{Y,5}$ (encoded in the sign of $c_5$) follows from the equations of chiral magnetohydrodynamics; see e.g.\ Ref.~\cite{Domcke:2019mnd}. 
In the following, the precise value of the   proportionality factor in Eq.~\eqref{eq:h} will be irrelevant, since, as we will see in the next section, the baryon asymmetry generated from this helicity is always far above the observed value once the CPI becomes fully efficient.

\subsection{Baryogenesis from decaying helical magnetic fields}
\label{subsec:YB_h}

There are two mechanisms for baryogenesis from the electroweak anomaly of the baryon symmetry in this setup. Firstly, as in spontaneous baryogenesis, the rotating axion can directly source a baryon asymmetry around the EWPT via the $SU(2)$ sphaleron process~\cite{Co:2019wyp}.%
\footnote{Earlier proposals in Refs.~\cite{Chiba:2003vp,Takahashi:2003db} also consider the production of the baryon asymmetry from the PQ charge, although they require processes that simultaneously break PQ and baryon symmetries, which are in fact not necessary as shown in Ref.~\cite{Co:2019wyp}.} 
However, if we require not to overproduce axion DM  via~\eqref{eq:DM}, this implies a baryon asymmetry below the observed value. Secondly, if the helical hypermagnetic fields survive until the EWPT (which is indeed the case; see App.~\ref{app:reynolds}), they can source a baryon asymmetry~\cite{Giovannini:1997gp,Giovannini:1997eg,Kamada:2016eeb,Kamada:2016cnb}. 
This can be understood from the different contributions of hypercharged gauge fields and electromagnetic gauge fields to anomalous violation of the baryon symmetry. In a SM-like crossover EWPT, this acts as a source term for the baryon asymmetry, while simultaneously weak sphaleron processes striving to wash out any $B+L$ asymmetry are becoming inefficient. The competition of these two processes determines the efficiency of this baryogenesis mechanism, referred to as baryogenesis from decaying helical magnetic fields.

We will focus on the latter mechanism here, which yields~\cite{Schober:2017cdw} 
\begin{align}
\label{eq:Bfromh}
Y_B \equiv \frac{n_B}{s} = c_B^\text{dec} \frac{\alpha_Y}{2 \pi} \frac{h}{s}  \,,
\end{align}
with $c_B^\text{dec} \simeq 0.05$ parametrizing the efficiency of baryogenesis from decaying helical magnetic fields at the EWPT~\cite{Kamada:2020bmb,Domcke:2022kfs}. Given the uncertainties in the dynamics of the EWPT, the possible value of $c_B^\text{dec}$ spans up to three orders of magnitude~\cite{Kamada:2016cnb,Jimenez:2017cdr}.

Let us first consider the case where the CPI becomes fully effective, $\Gamma_{\rm CPI} > \mathcal{O}(10) H$. We first note that, in this case, the baryon asymmetry is proportional to $\dot\theta / T$ based on Eqs.~\eqref{eq:h} and \eqref{eq:Bfromh}. Moreover, the produced helicity density $h$ does not get washed out, implying that the final baryon asymmetry is determined by the maximum value of $\dot\theta / T$ achieved during the entire cosmological evolution. As can be seen from the scaling of $\dot\theta$ in Eqs.~(\ref{eq:BG>_4})--(\ref{eq:BG<}), this maximum value occurs when $T = T_S$ and is given by
\begin{align}
\label{eq:max_dthetaOverT}
  \text{max}(|\dot \theta| /T ) = \frac{N_{\rm DW}m_\sx}{T_S} 
  \simeq 0.042 \, 
  c_\Omega^{-1/3} 
  \left( \frac{10^9~\text{GeV}}{f_a} \right)^{\scalebox{1.01}{$\frac{1}{3}$}} 
  \left( \frac{N_{\rm DW}\,m_\sx}{10^5~\text{GeV}} \right)^{\scalebox{1.01}{$\frac{2}{3}$}} 
  \left( \frac{g_*(T_S)}{g_{\rm MSSM}} \right)^{\scalebox{1.01}{$\frac{1}{3}$}}
  \,,
\end{align}
for both the quartic and quadratic potential. For the quartic case, $\dot \theta/T$ takes this constant value before $T_S$ and decreases after $T_S$, whereas $\dot\theta / T$ peaks at $T_S$ for the quadratic case.

Using Eqs.~\eqref{eq:h}--\eqref{eq:max_dthetaOverT}, we find the baryon asymmetry today
\begin{align}
\label{eq:YB_fullCPI}
\frac{|n_B|}{s} \sim 10^{-5}  \, c_\Omega^{-1/3} \left(\frac{c_B^\text{dec}}{0.05}\right) 
\left( \frac{N_{\rm DW}\,m_\sx}{10^5~\text{GeV}} \right)^{\scalebox{1.01}{$\frac{2}{3}$}}  
\left( \frac{10^9~\text{GeV}}{f_a} \right)^{\scalebox{1.01}{$\frac{1}{3}$}} 
\left( \frac{g_{\rm SM}}{g_*(T_{\rm EW})} \right)^{\scalebox{1.01}{$\frac{2}{3}$}} \,,
\end{align}
where $T_{\rm EW}$ is the temperature at the EWPT and $\text{sign}(n_B) = \text{sign}(c_5 \, \dot \theta)$. This overproduces the observed asymmetry of $Y_B^\text{obs} = 8.7 \cdot 10^{-11}$ from {\it Planck}~\cite{Aghanim:2018eyx} for
\begin{align}
 m_\sx \gtrsim  1 \MeV \, c_\Omega^{1/2} N_{\rm DW}^{-1} \, \left(\frac{0.05}{c_B^\text{dec}}\right)^{\scalebox{1.01}{$\frac{3}{2}$}} 
 \left( \frac{f_a}{10^9~\text{GeV}} \right)^{\scalebox{1.01}{$\frac{1}{2}$}}
 \left( \frac{g_*(T_{\rm EW})}{g_{\rm SM}} \right) \,, 
 \label{eq:mS_BAU_bound}
\end{align}
which in particular holds in the entire parameter regime where the CPI occurs based on Eqs.~\eqref{eq:mS_CPI} and \eqref{eq:mS_CPI_K}. Due to the overproduction by a large margin, this statement is robust against the large uncertainty in the efficiency of baryogenesis from decaying magnetic fields, i.e., the precise value of $c_B^\text{dec}$.

The observed amount of the baryon asymmetry can be obtained if the CPI becomes only marginally effective. This is possible if $\Gamma_{\rm CPI} = c_\text{CPI} H$ around $T_S$, since $\Gamma_{\rm CPI}/H$ is peaked at $T_S$ and decreases at $T<T_{S}$.

We now have all the ingredients to estimate the numerical value of $c_\text{CPI}$, as defined below Eq.~\eqref{eq:GammaCPI}.
The CPI enhances the magnetic field sourced by thermal fluctuations. The thermal number density of a helicity mode with a wave number $k_\text{CPI}$ is
\begin{align}
    h_{\rm th} \simeq \int_{k\simeq k_\text{CPI}} \frac{d^3 k}{(2\pi)^3}\frac{1}{e^{k/T}-1} \simeq \int_{k\simeq k_\text{CPI}} \frac{d^3 k}{(2\pi)^3}\frac{T}{k} \simeq \frac{1}{4\pi^2} k_\text{CPI}^2 T.
\end{align}
These thermal fluctuations are exponentially enhanced through the CPI, 
\begin{align}
\label{eq:h_exp}
    h = h_{\rm th} e^{\Gamma_{\rm CPI} t},
\end{align}
which then most efficiently sources the baryon asymmetry around $T \simeq T_S$. Using Eq.~\eqref{eq:Bfromh}, we can estimate the factor $c_\text{CPI}$,
indicating when these thermal fluctuations have grown to the level of the observed value of the baryon asymmetry,
\begin{align}
\label{eq:cCPI}
    c_\text{CPI} \simeq \frac{1}{Ht} \left( 11 + \frac{2}{3} \ln \left[
    \left( \frac{c_{\Omega} f_a}{10^6 \GeV} \right) 
    \left(\frac{10^4~{\rm GeV}}{N_{\rm DW}\,m_\sx}\right)^2 \left(\frac{1}{|c_5|}\right)^3
    \left(\frac{0.05}{c_B^{\rm dec}}\right)^{\scalebox{1.01}{$\frac{3}{2}$}}
    \left(\frac{g_*(T_S)}{g_{\rm MSSM}}\right)^{\scalebox{1.01}{$\frac{1}{2}$}}
    \right] \right) \,,
\end{align}
where $1/(Ht) =$ 2, 3/2, and 3 for RD, MD, and KD.
Note that in this estimate, we have neglected all non-linear terms in the magnetohydrodynamics equations, in particular those describing the effects of the plasma velocity, as well as backreaction effects which become relevant for large magnetic fields.  For numerical simulations including these terms, see Ref.~\cite{Schober:2017cdw}. 
Within this theoretical uncertainty, a value of $m_\sx$ saturating Eqs.~(\ref{eq:mS_CPI}) or (\ref{eq:mS_CPI_K}) with this value of $c_\text{CPI}$ can explain the observed baryon asymmetry.

\section{Results}
\label{sec:results}

\begin{figure}[t]
 \centering
 \includegraphics[width = 0.495 \textwidth]{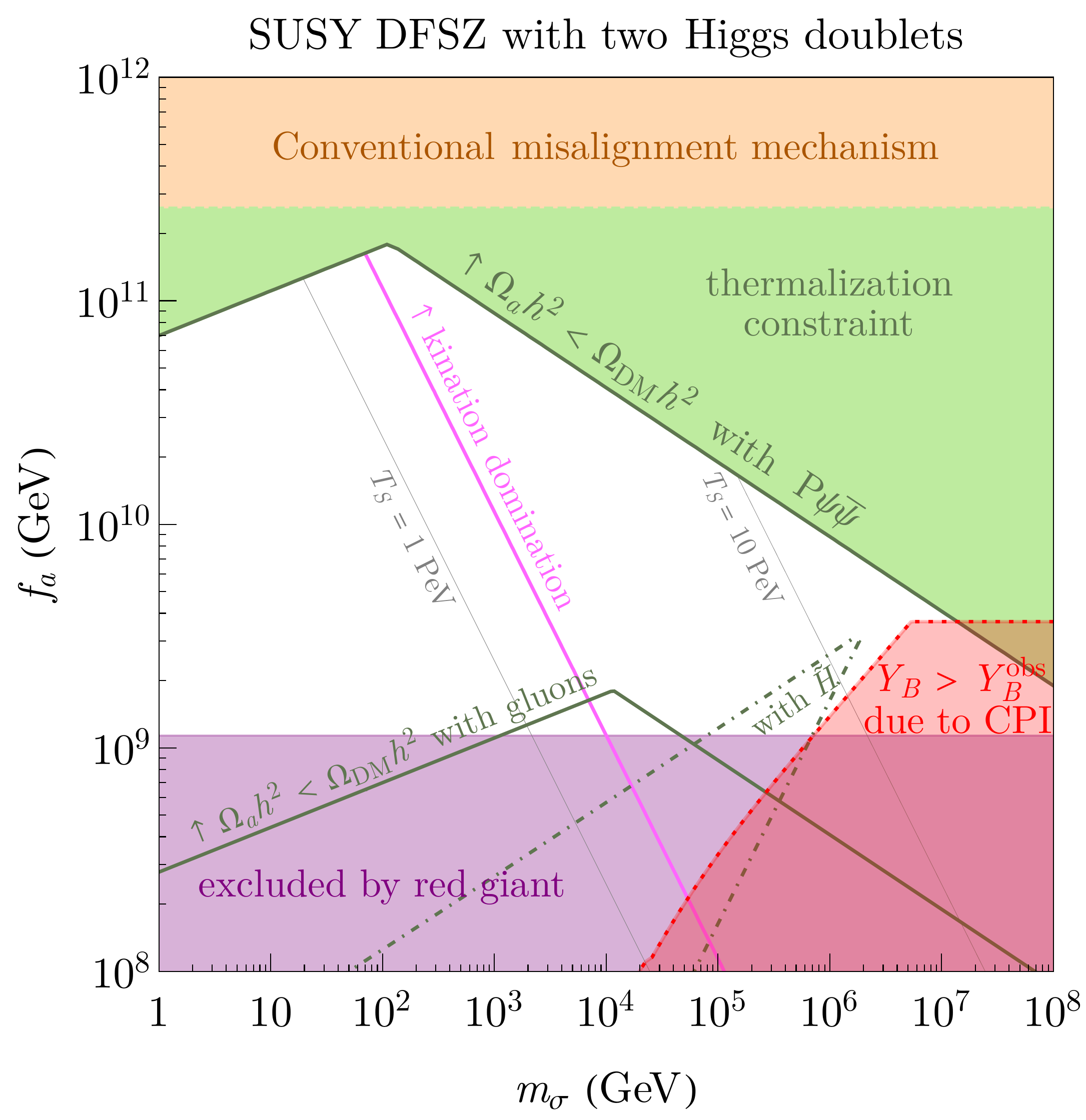} 
 \includegraphics[width = 0.495 \textwidth]{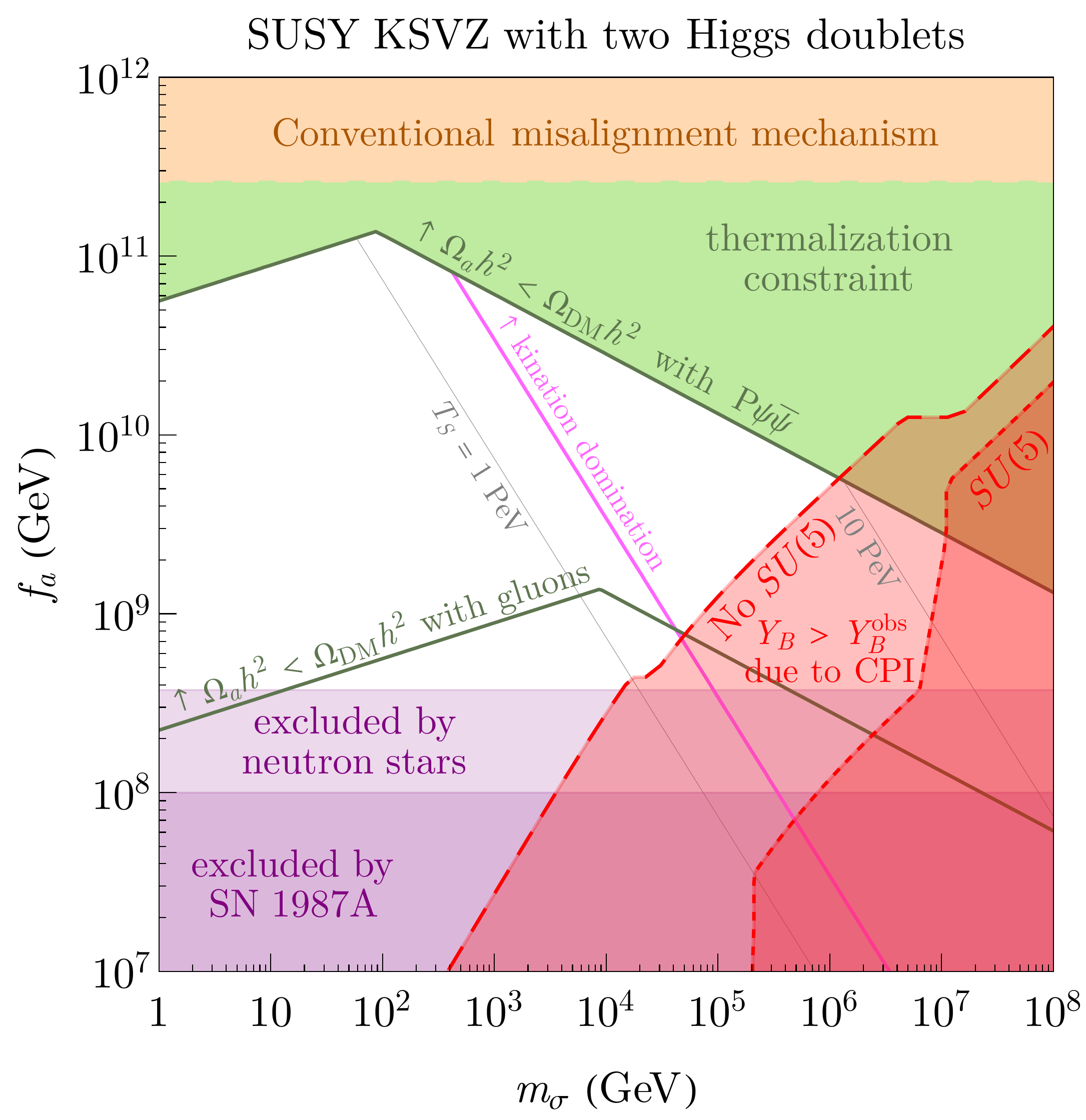}
 \caption{\footnotesize{Viable parameter space for the DFSZ (KSVZ) model in the left (right) panel. Here we consider supersymmetric models with two Higgs doublets. The shaded regions are excluded, see text, with the main new result of this work indicated by the red shaded region which leads to an overproduction of the baryon asymmetry. On the boundary of this region, the observed baryon asymmetry can be obtained. For the model parameter choices, see caption of Fig.~\ref{fig:c_5}.}} 
 \label{fig:SUSY_TwoHiggs}
\end{figure}

\begin{figure}[t]
 \centering
 \includegraphics[width = 0.495 \textwidth]{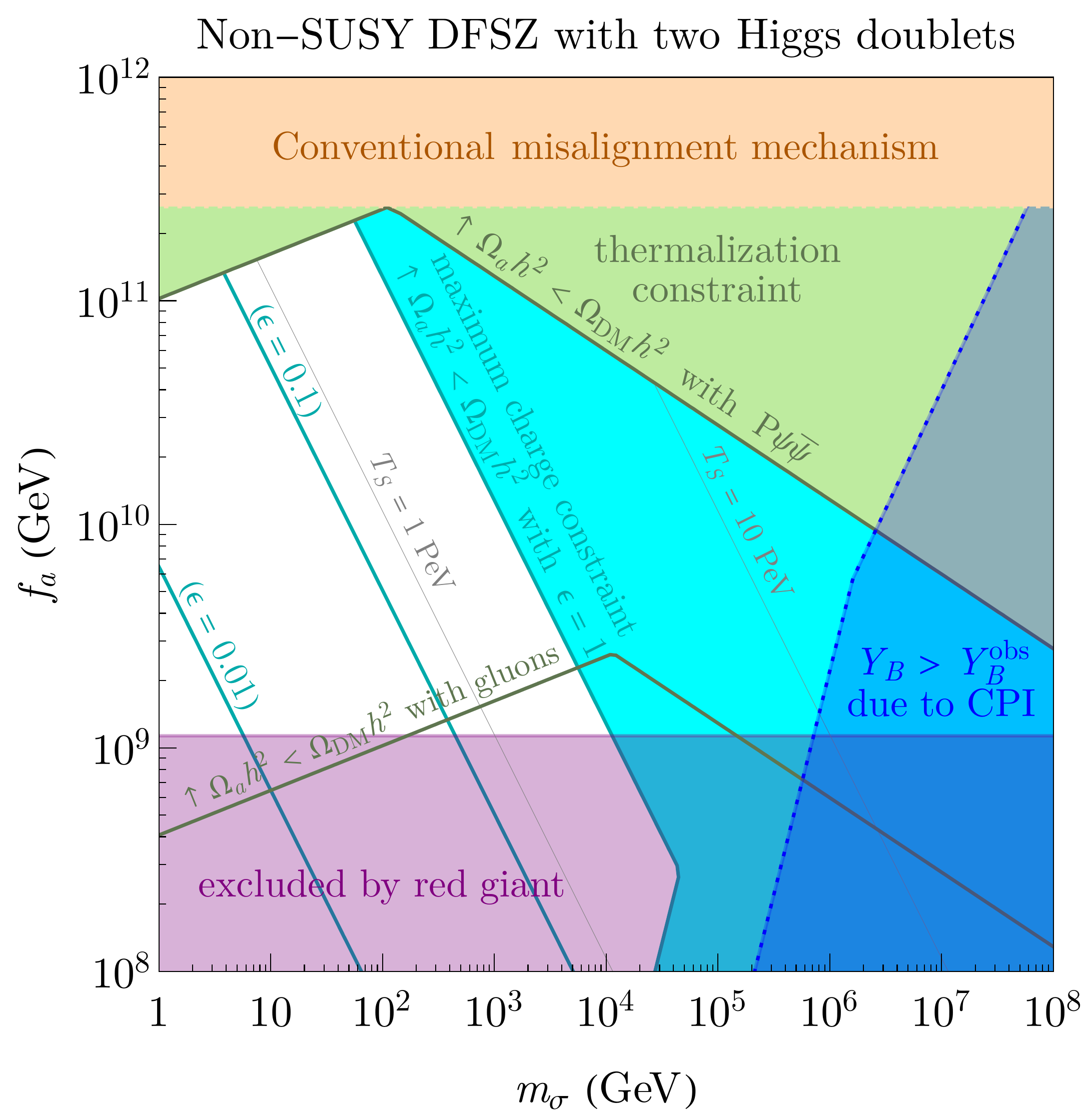} 
 \includegraphics[width = 0.495 \textwidth]{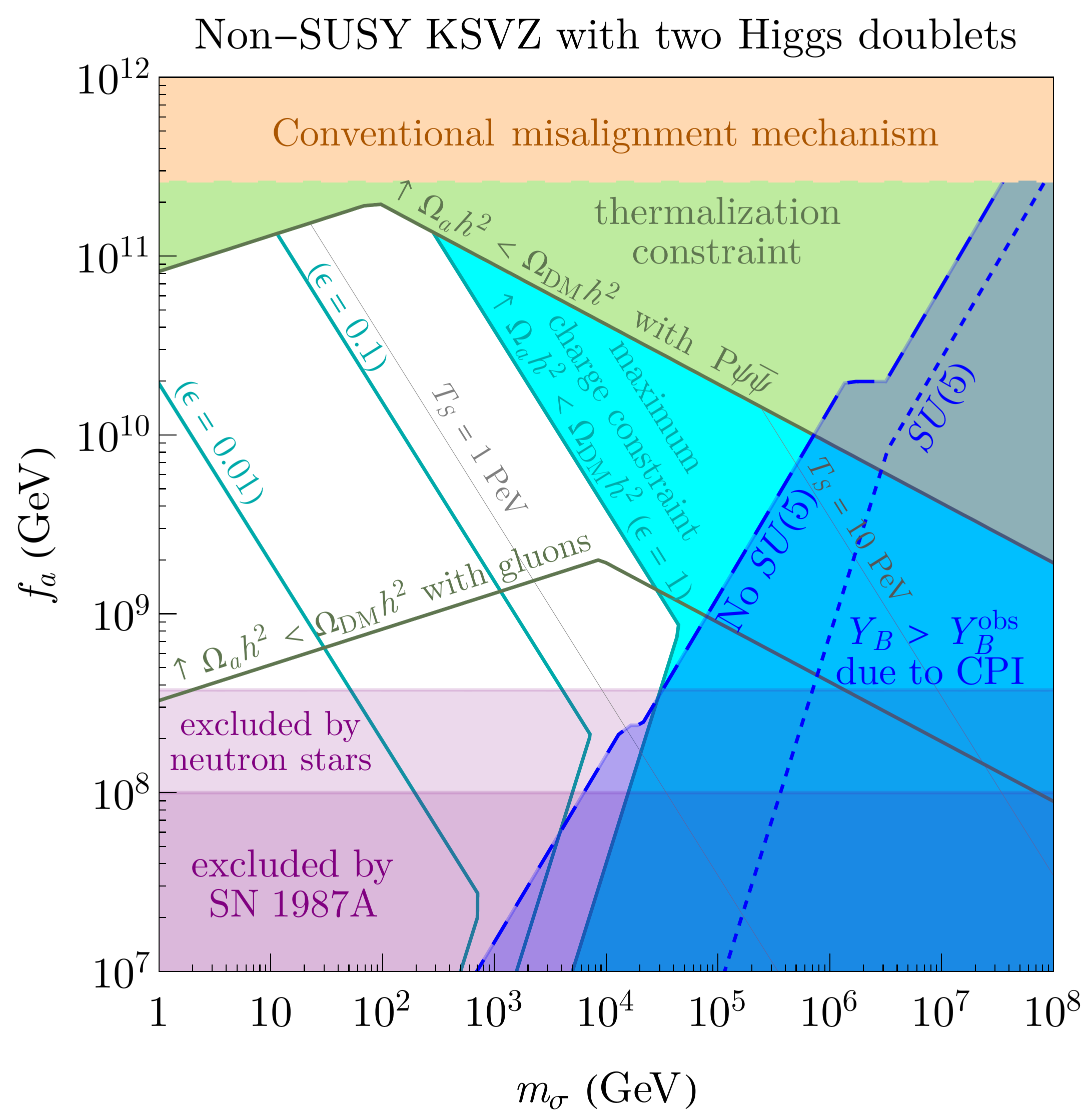}
 \caption{\footnotesize{Viable parameter space for the DFSZ (KSVZ) model in the left (right) panel. Here we consider non-supersymmetric models with two Higgs doublets. The shaded regions are excluded, see text, with the main new result of this work indicated by the blue shaded region which leads to an overproduction of the baryon asymmetry. For the model parameter choices, see caption of Fig.~\ref{fig:c_5}.}}
 \label{fig:NonSUSY_TwoHiggs}
\end{figure}

\begin{figure}[t]
 \centering
 \includegraphics[width = 0.495 \textwidth]{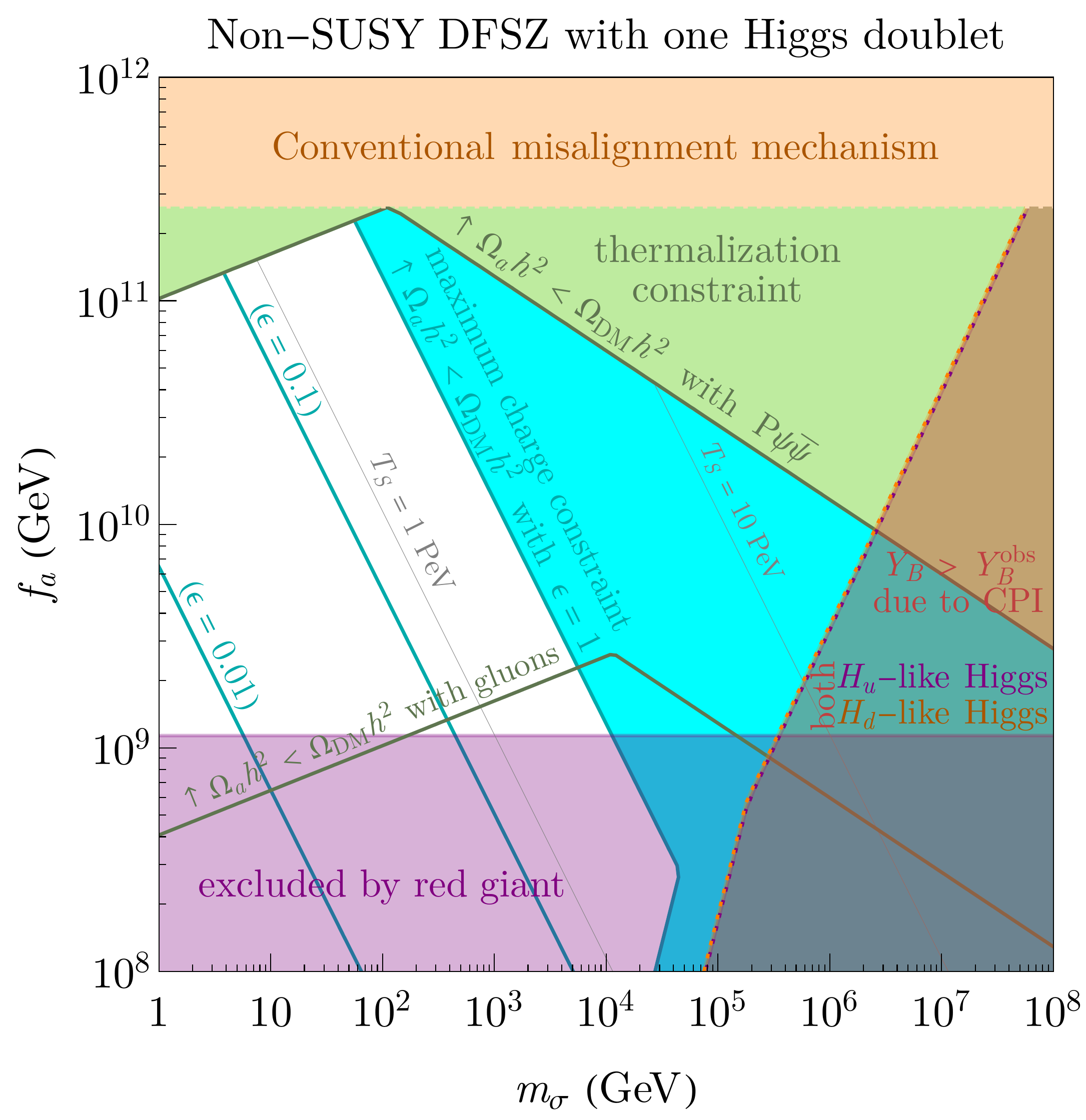} 
 \includegraphics[width = 0.495 \textwidth]{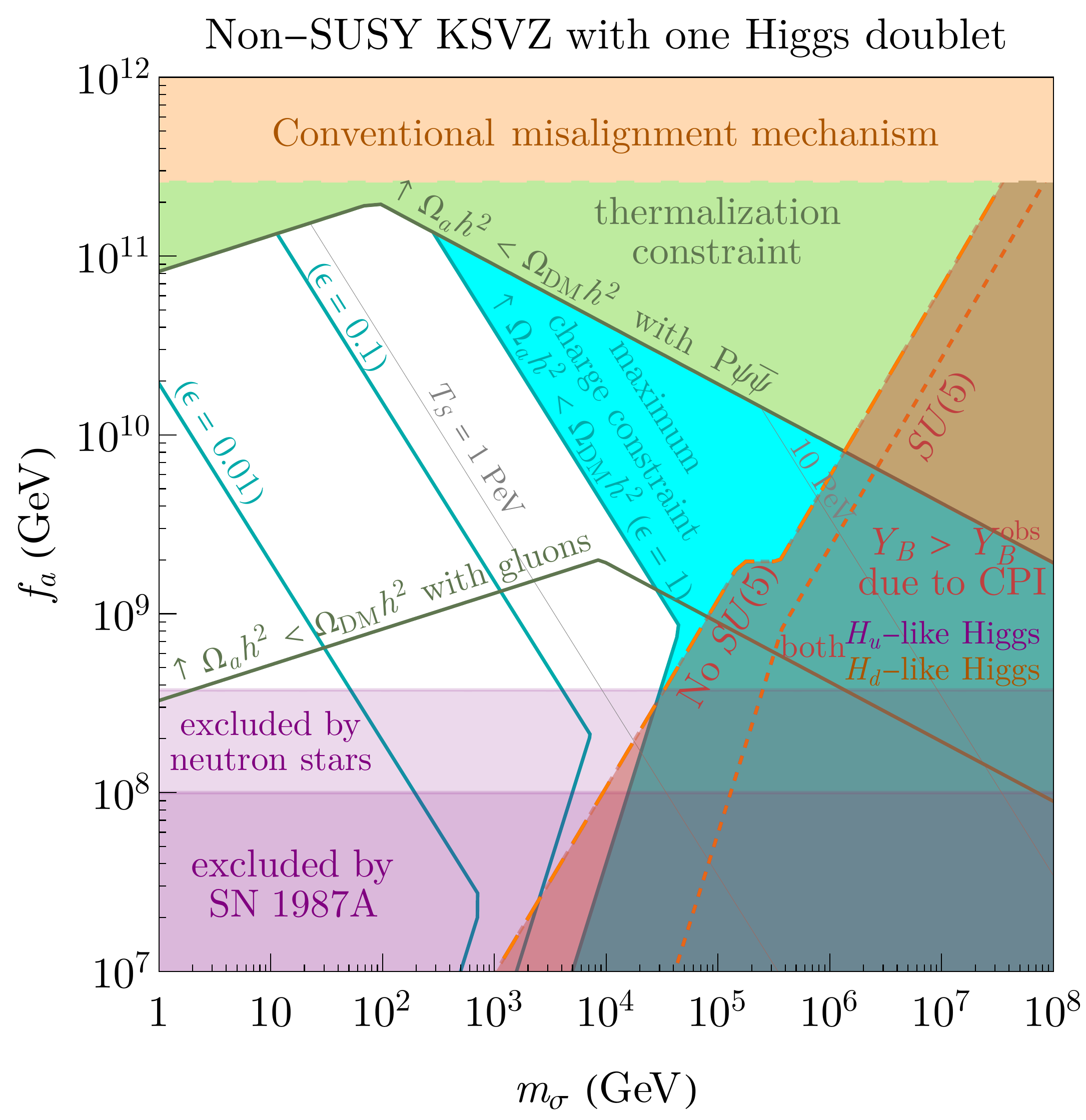}
 \caption{\footnotesize{Viable parameter space for the DFSZ (KSVZ) model in the left (right) panel. Here we consider non-supersymmetric models with one Higgs doublets. The shaded regions are excluded, see text, with the main new result of this work indicated by the maroon shaded region which leads to an overproduction of the baryon asymmetry. For the model parameter choices, see caption of Fig.~\ref{fig:c_5}.}} 
 \label{fig:OneHiggs}
\end{figure}

The (over)production of baryon asymmetry through the CPI leads to a new constraint on the parameter space of axion dark matter production through the kinetic misalignment mechanism, cf.\ Eqs.~\eqref{eq:mS_CPI} and \eqref{eq:mS_CPI_K}. This constraint sensitively depends on $c_5 = \mu_{Y,5}/\dot \theta$, and hence on the particle content and interactions in the thermal plasma. In Figs.~\ref{fig:SUSY_TwoHiggs}, \ref{fig:NonSUSY_TwoHiggs}, and \ref{fig:OneHiggs} we show detailed results for the parameter space of some of the most commonly discussed scenarios. 
The left (right) panel in each figure is for the DFSZ (KSVZ) theories. Inside the (red, blue, maroon) region labeled $Y_B > Y_B^\text{obs}$ of Figs.~(\ref{fig:SUSY_TwoHiggs}, \ref{fig:NonSUSY_TwoHiggs}, \ref{fig:OneHiggs}), the CPI occurs for the (SUSY two-Higgs, non-SUSY two-Higgs, non-SUSY one-Higgs) case, respectively. Here, the colors and boundary line styles follow those in Fig.~\ref{fig:c_5} for the corresponding $c_5(T)$, whereas the maroon shading in Fig.~\ref{fig:OneHiggs} is a result of overlaying purple and orange since $c_5(T)$ is identical for both the $H_u$-like and $H_d$-like Higgs scenarios in the parameter space shown. 
Inside all these shaded regions, the CPI is fully effective and the baryon asymmetry is overproduced according to Eq.~(\ref{eq:YB_fullCPI}). Along the boundaries of these regions, the CPI is only partially effective and the correct baryon abundance can be obtained for suitably tuned parameters. This can be understood as follows. For the cases where $\Gamma_{\rm CPI}/H$ peaks at $T_S$, $m_\sx$ needs to precisely follow Eqs.~(\ref{eq:mS_CPI}) and (\ref{eq:mS_CPI_K}) with $c_\text{CPI}$ given in Eq.~(\ref{eq:cCPI}). Tuning is necessary because, based on Eq.~(\ref{eq:h_exp}), the final baryon asymmetry is exponentially sensitive to this value of $m_\sx$. Since $c_5(T)$ changes with temperature, $\Gamma_{\rm CPI}/H$ does not necessarily peak at $T_S$. (For reference, we show contours of $T_S$ in the gray lines.) Therefore, to set the boundaries in the figures, we instead directly use Eqs.~(\ref{eq:Bfromh}) and (\ref{eq:h_exp}), though the results from these two prescriptions are nearly identical. As a main result of this work, these regions show that the CPI excludes the high-$m_\sx$ regions, assuming the kinetic misalignment origin of axion dark matter, due to the overproduction of the baryon asymmetry from the decaying magnetic helicity, while the values of $m_\sx$ along the boundaries give the correct baryon asymmetry.%
\footnote{In this calculation, we adopt the approximation that the scaling laws of $\dot\theta$ and $\rho_\theta$ abruptly change at $T_S$, while the actual transition is smooth for realistic potentials as discussed at the end of Appendix~\ref{app:scaling}. We show the effects of such smooth transitions in Fig.~\ref{fig:const_c5} by numerical calculations.}

In the quadratic case in Fig.~\ref{fig:SUSY_TwoHiggs}, explaining the observed baryon asymmetry from the CPI requires $m_\sx$ to be $\mathcal{O}(1 - 10)$ PeV  for both the DFSZ and the $SU(5)$-complete KSVZ models as shown by the red regions enclosed by the dotted and short-dashed lines. The mass scale can become as low as $\mathcal{O}(10)$ TeV for the KSVZ model without $SU(5)$ gauge unification.

In the quartic cases shown in Figs.~\ref{fig:NonSUSY_TwoHiggs} and \ref{fig:OneHiggs}, there already exists a strong constraint on high $m_\sx$ given in Eqs.~(\ref{eq:quartic_Ymax}) and (\ref{eq:quartic_Ymax_2}) from the maximum charge yield, and this constraint is shown as the cyan region in the figures for $\epsilon = 1$, corresponding to an initial motion that is circular. Other cyan lines assume more realistic $\epsilon = 0.1,~0.01$. We note that the positively-sloped cyan boundary comes from Eq.~(\ref{eq:quartic_Ymax_2}), where the maximum values of $H_I$ and $T_R$ allowed by CMB observations are assumed. Smaller values of $H_I$ and $T_R$ will make this segment of the cyan constraint more stringent, forbidding sufficient baryon asymmetry production following the CPI. The new constraint from the CPI does not impact the parameter space for the parameter choices shown here.

We now discuss the constraints from the green regions or potentially the green lines. The radial mode of $P$, denoted by $\sx$, initially carries a large energy density since the initial motion is not perfectly circular. Therefore, $\sx$ must be thermalized in order not to cause moduli problems. The thermalization constraints, however, depend on the particle content of the model. Here we summarize the results derived in Ref.~\cite{Co:2019jts} for the following two cases. In the minimal scenario where only the coupling of $P$ with gluons is present, the constraint is shown by the lower dark green line in each figure, labeled ``$\Omega_a h^2 < \Omega_\text{DM} h^2$ with gluons.'' Above these dark green lines, large $f_a$ suppresses the interaction rate to a point that thermalization occurs at a temperature too low to reproduce sufficient dark matter from kinetic misalignment. If an additional coupling $P\psi\bar\psi$ with fermions $\psi$ and $\bar\psi$ is introduced, the constraint can be relaxed to as far as the boundary of the green region. In these green lines/boundaries, the positively-sloped segments correspond to thermalization via perturbative decays, while the negatively-sloped segments are via thermal scattering. Lastly, a special thermalization channel~\cite{Barnes:2022ren} arises for DFSZ supersymmetric theories via the $\mu$ term as a coupling between $P$ and the Higgsinos, $\mu \sx^{n-1}/(N_{\rm DW} f_a)^n \times \sx \tilde{H}_u \tilde{H}_d$, from the superpotential $W = \mu P^n H_u H_d /(N_{\rm DW} f_a/\sqrt{2})^n$ that gives rise to the MSSM $\mu$ term. For $n = 1$ and $\mu = m_\sx$, the thermalization constraint is the green dot-dashed line in the left panel of Fig.~\ref{fig:SUSY_TwoHiggs}. For smaller $\mu/m_\sx$ ratios, this allowed region enclosed by the dot-dashed line would shift to the right. The constraints corresponding to the left and right dot-dashed lines are respectively given by
\begin{align}
    m_\sx & > 440 \TeV 
    \left( \frac{f_a}{2 \times 10^9 \GeV} \right)^3
    \left( \frac{g_*(T_{\rm th})}{g_{\rm MSSM}} \right)^{\scalebox{1.01}{$\frac{1}{2}$}} 
    \left( \frac{1}{\mu/m_\sx}\right)^2, \\
    m_\sx & < 1.2 \PeV
    \left( \frac{f_a}{2 \times 10^9 \GeV} \right)
    \left( \frac{g_{\rm MSSM}}{g_*(T_{\rm th})} \right)^{\scalebox{1.01}{$\frac{1}{6}$}}
    \left( \frac{1}{\mu/m_\sx}\right)^{\scalebox{1.01}{$\frac{4}{3}$}} .
\end{align}
For minimal DFSZ models, we note that the $\sx$-gluon coupling is only generated at temperatures below the Higgs masses. Thus, for the SUSY model in the left panel of Fig.~\ref{fig:SUSY_TwoHiggs}, only the green dot-dashed lines apply unless additional PQ-charged fermions are introduced, whether in the bath or integrated out to give a $\sx$-gluon coupling, to relax the constraint to the aforementioned green region or solid line. For the non-SUSY models in the left panels of Figs.~\ref{fig:NonSUSY_TwoHiggs} and \ref{fig:OneHiggs}, the thermalization constraint sensitively depends on the form of the Higgs coupling to $\sx$ and on the Higgs masses. Instead of exploring different models, we show again the constraints for the couplings with gluons (from integrating out heavy PQ charged fermions) and with new PQ-charged fermions in the bath. 

Finally, we briefly summarize the remaining constraints shown in Figs.~\ref{fig:SUSY_TwoHiggs}, \ref{fig:NonSUSY_TwoHiggs}, and \ref{fig:OneHiggs}. The purple regions are excluded by red giant brightness observations via the constraint on the axion-electron coupling~\cite{Capozzi:2020cbu,Straniero:2020iyi} for DFSZ (left panels) and by neutron star cooling~\cite{Iwamoto:1984ir, Iwamoto:1992jp, Page:2010aw, Shternin:2010qi, Leinson:2014cja, Leinson:2014ioa, Sedrakian:2015krq, Hamaguchi:2018oqw, Leinson:2021ety, Buschmann:2021juv} and SN 1987A cooling~\cite{Ellis:1987pk,Raffelt:1987yt,Turner:1987by,Mayle:1987as,Raffelt:2006cw,Chang:2018rso,Carenza:2019pxu} via on the axion-nucleon coupling for KSVZ (right panels). Lastly, we assume that the axion dark matter abundance comes from kinetic misalignment, but in the orange regions, an inconsistency in the calculations arises because the axion rotation fails to enhance the abundance, and conventional misalignment is operative instead~\cite{Co:2019wyp}. In this case, the CPI could still generate large helical magnetic fields at temperatures well above the QCD scale if the axion is initially rotating before becoming frozen by the Hubble friction. This case is however beyond the focus of the current work.

Before moving on to our conclusions, let us compare the constraint on $m_\sx$ obtained here with the parameter space obtained when using the dimension-five Majorana mass term of neutrinos to convert the charge of the axion rotation into a baryon asymmetry, referred to as  lepto-axiogenesis~\cite{Co:2020jtv}. There, for supersymmetric KSVZ/DFSZ models, $m_\sx$ is required to be $\mathcal{O}(10^3/10^2)$ TeV or below~\cite{Barnes:2022ren}, so lepto-axiogenesis is typically more efficient except for non-$SU(5)$ KSVZ models. In non-supersymmetric models, the required $m_\sx$ is about 6 times larger, and baryogenesis through the CPI can be more efficient for non-$SU(5)$ KSVZ models.
Finally, we note that the CPI can be more effective in extensions of DFSZ and KSVZ models. For example, the domain wall number $N_{\rm DW}$ in DFSZ models can be smaller if the PQ-symmetry breaking field couples to extra heavy quarks. Then the upper bound on $m_\sx$ becomes stronger as $N_{\rm DW}^2$ for the case with kination domination.

\section{Summary and discussion}
\label{sec:conclusions}

An axion rotating in field space generates helical hypermagnetic fields via the chiral plasma instability, which subsequently generate a baryon asymmetry in their decay across the electroweak phase transition. When the CPI is efficient, the helical density reaches a quasi-equilibrium value, which then leads to an overproduction of the baryon asymmetry. In particular in supersymmetric models this provides a stringent constraint in the paradigm where the axion serves as dark matter with the abundance generated by the rotation via kinetic misalignment. On the other hand, the observed baryon asymmetry can be generated if the CPI is only partially efficient, which is possible when the axion rotation slows down during the exponential growth of the helicity. Accordingly, the baryon asymmetry is exponentially sensitive to the parameters that determine the CPI rate and the time when the axion slows down, so a fine-tuning of parameters is necessary.

We focused on the QCD axion in two canonical models, the DFSZ and KSVZ model, where the axion couples to the Higgs doublets and gauge bosons, respectively. For each model, we also considered both supersymmetric and non-supersymmetric frameworks. Requiring the observed baryon asymmetry fixes the mass $m_\sx$ of the radial mode $\sx$ of the PQ-breaking field, while higher values of $m_\sx$ are excluded due to overproduction. 

The results for the supersymmetric cases are shown in Fig.~\ref{fig:SUSY_TwoHiggs}. The left panel is for the DFSZ model, where the baryon asymmetry can be explained by a SUSY scale $m_\sx = \mathcal{O}(1\mathchar`-20)$~PeV shown by the red dotted boundary. The right panel is for the KSVZ model, where the prediction is $m_\sx = \mathcal{O}(10\mathchar`-1000)$~TeV and $m_\sx = \mathcal{O}(10)$~PeV without and with $SU(5)$ gauge unification, respectively. We see that 10~TeV-scale supersymmetry is possible, an exciting prospect for future colliders. Also remarkably, the majority of the predictions point to high scale supersymmetry at $\mathcal{O}(100\mathchar`-1000)$ TeV, which is consistent with the so-called without-singlet scenario or mini-split supersymmetry~\cite{Giudice:1998xp,Wells:2003tf,ArkaniHamed:2004fb,Giudice:2004tc,Wells:2004di,Ibe:2006de,Acharya:2007rc,Hall:2011jd,Ibe:2011aa,Arvanitaki:2012ps,ArkaniHamed:2012gw}, which is well motivated by the ability to explain the observed Higgs mass, to predict the gauginos within the reach of the LHC and future colliders, and to resolve or relax problems involving the Polonyi field, the gravitino, as well as SUSY flavor or $CP$. 

The non-supersymmetric cases are shown in Figs.~\ref{fig:NonSUSY_TwoHiggs} and \ref{fig:OneHiggs} for two Higgs doublets and one Higss doublet, respectively. Since the potential of $\sx$ is taken to be quartic for non-SUSY models, the large initial potential energy already leads to strong constraints on the amount of PQ charge in the rotation shown as the cyan regions. For these non-supersymmetric models, the baryon asymmetry contribution via the CPI process is always insufficient in regions outside the cyan exclusion. \medskip

These results have consequences for other observables related to axion dark matter. 
The exclusion of higher $m_\sx$ by the overproduction of the baryon asymmetry can also be interpreted as a constraint on the duration of rotation energy domination. Previously, Refs.~\cite{Co:2021lkc,Gouttenoire:2021wzu,Gouttenoire:2021jhk} showed that the matter- and kination-dominated eras as a result of rotation domination lead to a modification of primordial gravitational wave spectra, serving as unique signatures of the axion rotation. For the QCD axion, the temperature at the transition from kination to radiation domination is predicted to be $T_{\rm KR} \simeq 1.7 \PeV$ from Eq.~(\ref{eq:TKR}). From Fig.~\ref{fig:NonSUSY_TwoHiggs}, one finds that $T_S \lesssim 25 \PeV$ in the left panel, and $T_S$ in this context is understood as the transition temperature from matter to kination domination, $T_{\rm MK}$. This corresponds to a constraint on $T_{\rm RM} \lesssim 10^{10} \GeV$.

Moreover, in all of the viable scenarios above, the decay constant $f_a$ is predicted to be below $10^{10}$~GeV. The QCD axion in this range may be probed by IAXO~\cite{Irastorza:2011gs,IAXO:2020wwp,IAXO:2019mpb}, BREAD~\cite{BREAD:2021tpx}, TOORAD~\cite{Marsh:2018dlj,Schutte-Engel:2021bqm}, ARIADNE~\cite{Arvanitaki:2014dfa,ARIADNE:2017tdd}, and LAMPOST~\cite{Baryakhtar:2018doz,Chiles:2021gxk}. \medskip

There exist many paths to expand the current work. The mechanism can be generalized to models with axion-fermion couplings. For example, mixing of KSVZ fermions with SM fermions can introduce axion-fermion couplings. We may then obtain $c_5 = \mathcal{O}(1)$, defined in Eq.~\eqref{eq:mu5}, even if the KSVZ fermions are in $SU(5)$-complete multiplets. The predictions with $c_5 = \mathcal{O}(1)$ will then be close to the CPI curves in Fig.~\ref{fig:const_c5} or the ``No $SU(5)$" curves in Figs.~\ref{fig:SUSY_TwoHiggs}, \ref{fig:NonSUSY_TwoHiggs}, and \ref{fig:OneHiggs}, which point to values of $m_\sx$ with more detection opportunities at future colliders than the original ``$SU(5)$" curves. Since the coupling with gluons is not essential, this analysis can also be straightforwardly generalized to generic rotating complex fields, such as scalar superpartners in Affleck-Dine scenarios and axion-like particles. The overproduction of baryon asymmetry through the CPI may thus also play a role e.g.\ in relaxion models~\cite{Graham:2015cka}.
Lastly, we have assumed an evolution of the axion rotation dynamics specific to the Affleck-Dine mechanism, where the rotation starts with a large radius $\sx$ that later redshifts to the minimum of the wine-bottle potential $V(\sx)$. The axion rotation may instead be initiated by a charge transfer from a different scalar field rotation~\cite{Domcke:2022wpb}. The distinctive features therein are that the axion rotation is always contained to the minimum of $V(\sx)$ and thus the axion velocity evolves differently, and that the asymmetries of fermions are also induced by the rotation of the other scalar field. Such features will affect the phenomenological implications.

As described above, the baryon asymmetry in the baryogenesis mechanism presented in this work is exponentially sensitive to the parameters of the theory, and fine-tuning of them is required.
Nevertheless, the mechanism works without modifying the electroweak phase transition, which was required in minimal axiogenesis, and therefore we consider it to be a minimal scenario. The fine-tuning problem can be avoided by taking $m_\sx$ above the required value, producing the helical magnetic field to quasi-equilibrium in Eq.~\eqref{eq:h}, initially overproducing baryon asymmetry, and diluting it to the observed value by subsequent entropy production. This also dilutes the axion abundance, but the correct abundance can be maintained by also initially overproducing the axions from kinetic misalignment. We leave the investigation of the viable parameter region to future work.

Finally, our work may also have implications for intergalactic magnetic fields, whose existence has been suggested by blazar observations~\cite{Tavecchio:2010mk,Neronov:2010gir}. The magnetic field strength necessary to explain such present-day blazar observations would lead to an overproduction of the baryon asymmetry during the EWPT~\cite{Kamada:2016cnb}. An interesting resolution arises in our current framework because the CPI may become efficient only after EWPT. This points to a drastically different parameter space from this work, e.g., a much lighter $\sx$ field and/or a much lighter axion-like particle. The exploration of this possibility is left for future work.

\section*{Acknowledgement}
The authors thank Kohei Kamada, Kyohei Mukaida, Kai Schmitz and Masaki Yamada for helpful discussions. The work of R.C.~was supported in part by DOE grant DE-SC0011842 at the University of Minnesota.

\appendix

\section{Scaling of PQ fields}
\label{app:scaling}

Consider a scalar field $\phi$ described by the equation of motion in an expanding universe,
 \begin{align}
  \ddot \phi + 3 H \dot \phi + \frac{\partial V}{\partial \phi^*} = 0 \,.
 \end{align}
 Multiplying this equation with $\dot \phi^* = d\phi^*/dt$, adding the hermitian conjugate, and using $\rho_\phi = |\dot \phi|^2 + V(\phi)$ this immediately yields
 \begin{align}
  \dot \rho_\phi + 6 H |\dot \phi|^2 = 0 \,.
 \end{align}
Using the virial theorem, $2 \langle |\dot \phi|^2 \rangle = p \langle V(\phi) \rangle$ for a scalar potential $V(\phi) \propto |\phi|^p$, we find
 \begin{align}
  \dot \rho_\phi + 3 H \frac{2 p}{p + 2} \rho_\phi = 0 \,.
 \end{align}
From this we can immediately see that the energy density in a scalar potential with $p=2$ redshifts as matter, $\dot \rho_\phi + 3 H \rho_\phi = 0$, whereas a scalar potential with $p=4$ redshifts as radiation, $\dot \rho_\phi + 4 H \rho_\phi = 0$.

Applying this to PQ sector and taking into account the conserved comoving charge $\dot \theta r^2 /T^3 =$ constant, we find the scaling behaviour given in Eqs.~\eqref{eq:BG>_4} and \eqref{eq:BG>_2}. The virial theorem also immediately gives the solution for $\dot \theta(r)$ in both cases.

\begin{figure}[t]
 \centering
 \includegraphics[width = 0.495 \textwidth]{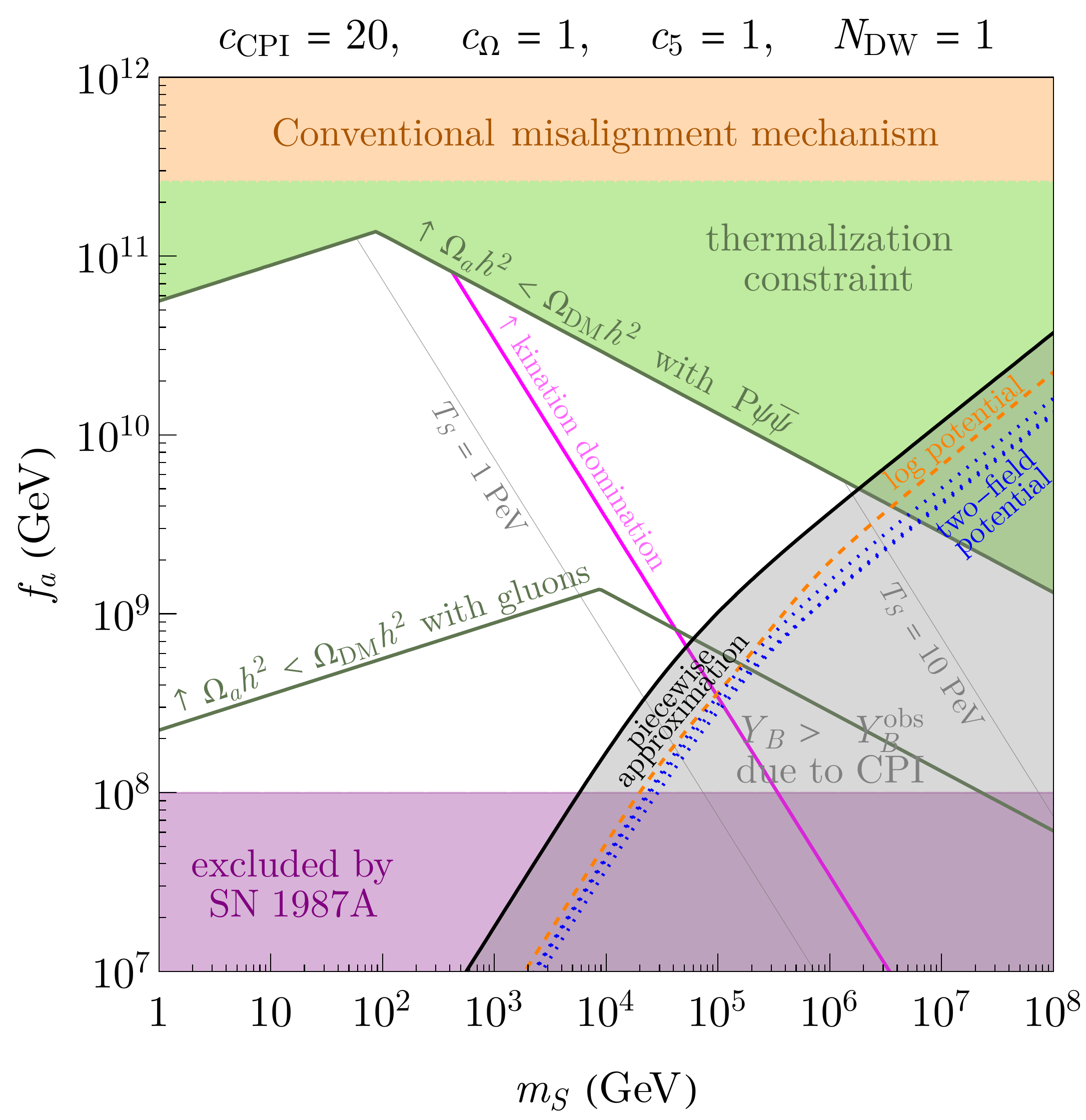} 
 \caption{\footnotesize{The comparison of the CPI constraints (gray shading) from different PQ-breaking potentials shown by the black, orange dashed, and blue dotted lines. The relevant dimensionless constants are specified at the top of the figure, while $c_5$ is fixed to unity in this figure simply for clarity.}} 
 \label{fig:const_c5}
\end{figure}

We note that the scaling laws derived above are valid when the potential is (locally) well approximated by $|\phi|^p$ with a constant value of $p$. In the transition regime when $r$ is settling down to the minimum of the potential, how the scaling of $\rho_\theta$ in Eqs.~\eqref{eq:BG>_4} and \eqref{eq:BG>_2} evolves towards Eq.~\eqref{eq:BG<} depends on the exact symmetry-breaking potential in the radial mode. This effect is analyzed in Ref.~\cite{Co:2021lkc} in the context of gravitational waves, but we will use the results of the energy density scaling. In Fig.~\ref{fig:const_c5}, we compare the constraints that lead to overproduction of the baryon asymmetry from the CPI, as discussed in Sec.~\ref{sec:BAU}, for different symmetry-breaking potentials. For illustration, we focus on the radial mode potential that is approximately quadratic in the large field values, which is the case for supersymmetric theories as discussed below Eq.~(\ref{eq:P}) and elaborated in Ref.~\cite{Co:2021lkc}. The gray shaded region applies if the rotation energy density scaling changes suddenly from matter to kination when $r$ settles to the minimum---a piecewise approximation that would not be realistic. Actual models, such as the log potential from dimension transmutation and the two-field model analyzed in Ref.~\cite{Co:2021lkc}, will lead to a smoother transition from matter to kination. This modifies the gray region such that the boundary shifts from the black line to the orange dashed line for the log potential. On the other hand, the boundary instead moves to the blue dotted lines for the two-field model, where the multiple lines show the variations when the ratio of the soft masses of the two fields in the model is varied over all possible values. As seen from Fig.~\ref{fig:const_c5}, the boundary may shift up to a factor of $\mathcal{O}(3)$ in $m_\sx$ between different models. For simplicity in presenting Figs.~\ref{fig:SUSY_TwoHiggs}, \ref{fig:NonSUSY_TwoHiggs}, and \ref{fig:OneHiggs} , we only show the results that adopt the piecewise approximation.

\section{Computation of $c_5$}
\label{app:c5}
In this Appendix, we compute the coefficient $c_5$ that relates the axion field velocity and the chiral chemical potential as defined in Eq.~\eqref{eq:mu5}.
To be concrete, we compute it in the SM, in the type-II two Higgs-doublet extension, and in the MSSM. We will refer to them as ``non-SUSY one Higgs", ``non-SUSY two Higgs", and ``SUSY", respectively.
In SUSY models, we assume all sparticles have masses far below $T_S$. We moreover consider the KSVZ and DFSZ axion models.    
We use the convention where all fermions fields are left-handed Weyl fields. In particular, $SU(2)_L$-singlet fields in the (MS)SM are written as $\bar{u}$, $\bar{d}$, and $\bar{e}$, and chemical potentials are defined for left-handed fermions. The correspondence with the notation in e.g.~\cite{Domcke:2020kcp} is $\mu_{\bar{u}} = - \mu_{u_R}$, etc.

We begin with the following coupling of the axion with the gauge bosons and Higgs fields,
\begin{align}
{\cal L} = \frac{\theta}{64\pi^2} \epsilon^{\mu \nu \rho \sigma} \left( c_gg_3^2 G_{\mu\nu}^a G_{\rho \sigma}^a + c_Wg_2^2 W_{\mu\nu}^i W_{\rho \sigma}^i + c_Yg_Y^2 B_{\mu\nu} B_{\rho \sigma} \right)  + B \mu e^{i n\theta/N_{\rm DW}} H_u H_d,
\label{eq:Lapp}
\end{align}
where $\theta = a / f_a$ is the canonically normalized axion field, $N_{\rm DW}$ is the domain wall number, and $n$ is an integer. For the KSVZ model, $c_g =1$ and $n=0$, and $c_W$ and $c_Y$ depend on the gauge charge of the KSVZ fermions. For example, for a $SU(5)$-complete KSVZ fermion, $c_W=1$ and $c_Y=5/3$.  For the DFSZ model, $c_g=c_W=c_Y=0$ and $N_{\rm DW} = 3n$. In the MSSM, these couplings arise from the coupling of the axion through the gauge kinetic terms and the superpotential $W \propto P^n H_u H_d$.
In non-SUSY models, it is possible that one linear combination of $H_u$ and $H_d$ is heavy. After integrating out the heavy combination, the axion couples to SM particles via the axion-dependent down-type Yukawa couplings $y_d e^{-i \theta/N_{\rm DW}}$ for an $H_u$-like SM Higgs and up-type Yukawa couplings $y_u e^{-i \theta/N_{\rm DW}}$ for an $H_d$-like SM Higgs.

As we discuss in the main text, we are interested in the temperature $T_S$, at which the radial direction reaches its local minimum. In the MSSM, the scattering by the gaugino mass is already in thermal equilibrium around this temperature, so we may take the chemical potential of the gauginos to be zero. The scalar and fermion components of chiral multiplets then have the same chemical potential. The relations among the chemical potentials, governed by the relevant Yukawa interactions and sphaleron processes, take similar forms in non-supersymmetric and supersymmetric theories.

The relations among the chemical potentials in general depend on the coupling constants. This is because the PQ symmetry has the QCD anomaly, and at the true equilibrium, $\dot{\theta}=0$ and $\mu_i=0$. In the cosmological context, the system may be in quasi-thermal equilibrium state where the axion rotational velocity takes a value determined by $\dot{\theta}\simeq N_{\rm DW} \sqrt{V'(\sx)/\sx}$ which is slowly washed out. To obtain the chemical potential of particles in this axion background, one must write down the relevant Boltzmann equations for all interactions, and set the time derivative of all chemical potentials to be zero. This in general leads to chemical potentials that non-trivially depend on the strength of these interactions, i.e., the coupling constants of the theory. A great simplification occurs, however, because of the small up Yukawa coupling. When we expand the equations for the chemical potential with respect to the up Yukawa coupling, the leading $\mathcal{O}(y_u^0)$ term coincides with the case where the up Yukawa is taken to be zero, for which the PQ symmetry is restored. We may then apply the detailed balance relation to each interaction, obtaining a set of algebraic equations.%
\footnote{
In the language of Ref.~\cite{Domcke:2020kcp}, this set of algebraic equations corresponds to taking all interactions either fully equilibrated or completely irrelevant, while neglecting the backreaction on the axion motion. As noted there, special care must be taken when interactions are not linearly independent. Using the tools of~\cite{Domcke:2020kcp}, one can easily reproduce the set of equations given below. The backreaction to the axion motion becomes relevant when the axion couples to an operator whose charge vector lies in the span of the charge vectors of the efficient interactions. In our case, this happens when 
the axion couples to $\epsilon^{\mu \nu \rho \sigma} G_{\mu \nu}^a G_{\rho \sigma}^a$ and 
the up Yukawa is efficient, since the up Yukawa interaction and the strong sphaleron process are not linearly independent.
}
As we will see, however, there are several cases where $c_5$ vanishes under this approximation. We will take into account the effect of non-zero up Yukawa at the end of this Appendix.

We first consider the case where all SM interactions, including the electron Yukawa as the weakest interaction, are in thermal equilibrium.
The relations between the chemical potentials are then given by
\begin{align}
&\text{strong sphaleron} \quad \quad  &&\sum_i\left(2\mu_{q_i} + \mu_{\bar{u}_i} + \mu_{\bar{d}_i}\right) + c_g \dot{\theta} = 0, \\
&\text{weak sphaleron} \quad 
&& \begin{cases}
\sum_i\left(3\mu_{q_i} + \mu_{\ell_i} \right) + c_W \dot{\theta} = 0, & \text{: non-SUSY} \\
\sum_i\left(3\mu_{q_i} + \mu_{\ell_i} \right) + \mu_{H_u} + \mu_{H_d} + c_W \dot{\theta} = 0, & \text{: SUSY}
\end{cases} \\
 \label{eq:mu}
& W \supset P^n H_u H_d \quad  && \mu_{H_u} + \mu_{H_d} - \frac{n}{N_{\rm DW}} \dot{\theta} = 0,  \\
\label{eq:ye}
& \text{lepton Yukawa} \quad    && \mu_{\bar{e}_i} + \mu_{\ell_i} + \mu_{H_d} = 0,  \\
\label{eq:yd}
& \text{down-quark Yukawa} \quad     && \mu_{\bar{d}_i} + \mu_{q_j} + \mu_{H_d} = 0, \\
& \text{up-quark Yukawa} \quad    && \mu_{\bar{u}_{2,3}} + \mu_{q_{2,3}} + \mu_{H_u} = 0.
\end{align}

These equations assume the existence of two Higgses $H_u$ and $H_d$ in the bath, which may not be the case for non-SUSY theories; it is possible that one of them does not exist in the KSVZ or is heavy even in the DFSZ.
For the KSVZ model, we may still continue to use the equations since Eq.~\eqref{eq:mu} enforces $\mu_{H_u} = - \mu_{H_d}$, and inserting this relation in the other equations, we recover the result for the case with only one Higgs $H$.
The required modification is slightly more complicated for the DFSZ theory with only one Higgs in the bath. After integrating out the heavy Higgs to obtain the effective theory including the SM and the axion, the Yukawa couplings depend on the axion, and the axion couples to the Higgs current. The former effect is encoded by Eq.~\eqref{eq:mu}, but the latter effect is missed. A simplification occurs when the light Higgs is dominantly $H_u$ or $H_d$, since the axion-Higgs coupling is suppressed. In the following, we present the result for the case where the SM Higgs is dominantly $H_u$ or $H_d$.

The conserved charges are hypercharge and the flavoured $B$$-$$L$ (baryon number minus lepton number) charges,
\begin{align}
    Y & \propto
    \begin{cases}
    \sum_i \left( \mu_{q_i} -2 \mu_{\bar{u}_i} + \mu_{\bar{d}_i} - \mu_{\ell_i} + \mu_{\bar{e}_i} \right) + \mu_{H_u} - \mu_{H_d} & \text{: SUSY} \\
    \sum_i \left( \mu_{q_i} -2 \mu_{\bar{u}_i} + \mu_{\bar{d}_i} - \mu_{\ell_i} + \mu_{\bar{e}_i} \right) + 2 (\mu_{H_u} - \mu_{H_d})  & \text{: non-SUSY, two Higgses } \\
       \sum_i \left( \mu_{q_i} -2 \mu_{\bar{u}_i} + \mu_{\bar{d}_i} - \mu_{\ell_i} + \mu_{\bar{e}_i} \right) + 2 \mu_{H_u}   & \text{: non-SUSY, $H_u$-like Higgs} \\
       \sum_i \left( \mu_{q_i} -2 \mu_{\bar{u}_i} + \mu_{\bar{d}_i} - \mu_{\ell_i} + \mu_{\bar{e}_i} \right) - 2 \mu_{H_d}   & \text{: non-SUSY, $H_d$-like Higgs}
    \end{cases} \nonumber \\
    \frac{B}{3} - L_i & \propto \frac{1}{3} \sum_j\left(2 \mu_{q_j} - \mu_{\bar{u}_j} -  \mu_{\bar{d}_j} \right) -2 \mu_{\ell_i} + \mu_{\bar{e}_i}\,.
\end{align}

Based on the above equations, we obtain\footnote{
The contribution proportional to $c_Y$ is obtained from Eq.~\eqref{eq:Lapp} by deriving the equation of motion of hypercharge gauge field to find $\nabla \times B_Y - \dot{E}_Y =  - (c_Y\alpha_Y\dot{\theta}/2\pi) B_Y$.}
\begin{align}
    c_5 = \mu_{Y,5}/ \dot{\theta} & = -\sum_i g_i Y_i^2 \frac{\mu_i}{\dot{\theta}} -  \frac{1}{2}  c_Y \nonumber  =
     \begin{cases}
     -\frac{n}{N_{\rm DW}} + \frac{4}{3}c_g  -  \frac{1}{2}c_W - \frac{1}{2} c_Y & \text{SUSY} \\
     0\times \frac{n}{N_{\rm DW}} + \frac{4}{3}c_g  -  \frac{1}{2}c_W - \frac{1}{2} c_Y& \text{non-SUSY} \,.
     \end{cases} 
\end{align}

We next consider the case where (only) the electron Yukawa is out of equilibrium. The condition for $i=1$ in Eq.~\eqref{eq:ye} should be replaced by a conservation of $\bar{e}_1$ number,
\begin{align}
  \mu_{\bar{e}_1} = 0.  
\end{align}
We then obtain
\begin{align}
    c_5 
     & =
     \begin{cases}
     -\frac{460}{481}\frac{n}{N_{\rm DW}} + \frac{121}{111}c_g  -  \frac{349}{962}c_W - \frac{1}{2} c_Y & \text{SUSY} \\
     -\frac{90}{563}\frac{n}{N_{\rm DW}} + \frac{1874}{1689}c_g  -  \frac{40 7}{1126}c_W - \frac{1}{2} c_Y& \text{non-SUSY, two Higgses} \\
     -\frac{90}{481}\frac{n}{N_{\rm DW}} + \frac{121}{111}c_g  -  \frac{349}{962}c_W - \frac{1}{2} c_Y & \text{non-SUSY, $H_u$-like Higgs}
      \\
     0\times \frac{n}{N_{\rm DW}} + \frac{121}{111}c_g  -  \frac{349}{962}c_W - \frac{1}{2} c_Y & \text{non-SUSY, $H_d$-like Higgs} \,.
     \end{cases}
\end{align}

Finally, we consider the case where the down Yukawa is also out of equilibrium. The condition for $i=1$ in Eq.~\eqref{eq:yd} should be replaced by a conservation of $\bar{u}_1- \bar{d}_1$ number,
\begin{align}
  \mu_{\bar{u}_1} - \mu_{\bar{d}_1} = 0.  
\end{align}
We then obtain
\begin{align}
    c_5 
     & =
     \begin{cases}
     \frac{18}{179}\frac{n}{N_{\rm DW}} + \frac{389}{537}c_g  -  \frac{127}{358}c_W - \frac{1}{2} c_Y & \text{SUSY} \\
     \frac{21}{22}\frac{n}{N_{\rm DW}} + \frac{239}{330}c_g  -  \frac{39}{110}c_W - \frac{1}{2} c_Y& \text{non-SUSY, two Higgses} \\
     \frac{169}{179}\frac{n}{N_{\rm DW}} + \frac{389}{537}c_g  -  \frac{127}{358}c_W - \frac{1}{2} c_Y& \text{non-SUSY, $H_u$-like Higgs}  \\
     \frac{173}{179}\frac{n}{N_{\rm DW}} + \frac{389}{537}c_g  -  \frac{127}{358}c_W - \frac{1}{2} c_Y& \text{non-SUSY, $H_d$-like Higgs} \,.
     \end{cases}
\end{align}

In the KSVZ model, if the coupling of the axion with the gauge bosons satisfies the GUT relation, $c_5=0$ when the down Yukawa is in equilibrium. 
For such a case, the effect of non-zero up Yukawa can be important. To take this into that account, we can no longer use the detailed balance equation for each interaction. Taking the time derivative of the asymmetry of SM particles in the Boltzmann equation in Ref.~\cite{Co:2020xlh}, we obtain 
\begin{align}
    c_5 =
    \begin{cases}
     - \frac{y_u^2}{y_u^2 + y_d^2} c_g & y_e~\text{in equilibrium} \\
     - \frac{51y_d^2 + 397 y_u^2}{481(y_u^2 + y_d^2)} c_g & y_e~\text{out of  equilibrium} .
    \end{cases}
\end{align}
Here we take $n=0$ and $c_W= 3c_Y/5 = c_g= 1$ for the SUSY or non-SUSY one Higgs case; since we here consider the KSVZ model, we omit the non-SUSY two Higgs case. The effect is only an $\mathcal{O}(1)$ change in $c_5$  when the electron Yukawa is out of equilibrium, but when it is in equilibrium, $c_5$ is dominated by this effect.
In supersymmetric theories, however, $y_d$ is enhanced by ${\rm tan} \beta$, and $c_5$ remains much smaller than $\mathcal{O}(1)$.
We expect the deviation from the ideal gas approximation to give a further correction to $c_5$.
However, as discussed in the main text, the value of $c_5$ with $y_e$ in equilibrium affects the bound on $m_\sx$ only for~$f_a < 10^8$~GeV.

\section{Helicity density at equilibrium}
\label{app:helicity}

In this appendix, we estimate the helicity density when the CPI becomes fully effective and show that the charge in the axion rotation remains almost the same.  For simplicity, we consider a toy model of QED with a massless electron and a photon that couples to the axion. In this toy model, the instability is induced by the direct axion-photon coupling rather than by the fermion chiral asymmetry, but the toy model already shows the essence of why the axion rotation remains undisturbed. The system has two conserved charges,
\begin{align}
\label{eq:charges}
\dot{\theta} f_a^2 + \frac{\alpha}{2\pi} h \equiv q,\nonumber\\
q_e + \frac{\alpha}{2\pi} h = 0,
\end{align}
where $h$ is the helicity density, $q$ is the total charge, $q_e = (q_{e_L} - q_{e_R})/2$ is the chiral asymmetry of the electron, and $\alpha$ is the fine-structure constant.

The contributions to the free energy that depend on the axion rotation, the helicity density, and the electron chiral asymmetry are given by
\begin{align}
    F \supset \frac{1}{2}\dot{\theta}^2 f_a^2 + k_h |h| + \frac{6q_e^2}{T^2} ,
\end{align}
where $k_h$ is the typical wavenumber of the helical photons. The free energy decreases with decreasing $k_h$~\cite{Joyce:1997uy}, and the inverse cascade indeed drives $k_h$ to small values~\cite{Brandenburg:2017rcb,Schober:2017cdw,Schober:2018ojn}, so we drop the second term. Eliminating $q_e$ and $\dot{\theta}$ in favor of $h$ and ignoring the terms suppressed by $T^2/f_a^2$, we obtain
\begin{align}
    F \supset 6 \left(\frac{\alpha}{2\pi}\right)^2 \frac{h^2}{T^2} - \frac{1}{f_a^2} \frac{\alpha}{2\pi} h \, q.
\end{align}
The free energy is thus minimized for
\begin{align}
    \frac{\alpha}{2\pi} h \simeq \frac{1} {12} \frac{T^2}{f_a^2}q.
    \label{eq:happ}
\end{align}
This is much smaller than the total charge $q$ as long as $f_a \gg T$, so most of the charge remains stored in the axion rotation even when the CPI is efficient. The electron chiral asymmetry is as small as $q \, T^2/f_a^2$, which only makes up a small fraction $(T/f_a)^2$ of the charge initially stored in the axion rotation.

The reason why the PQ charge is dominantly stored in the axion rotation is the second conserved charge in Eq.~\eqref{eq:charges}. Indeed, if only the first one is conserved, the free energy of the system would be minimized when all of the charge in the axion rotation is transferred into the helicity density, since the energy per charge (i.e., chemical potential) of the helical photon can be much smaller than that of the axion rotation if $k_h \ll \dot{\theta}\alpha/\pi$. With the second conservation law, a large helicity density requires a large electron chiral asymmetry, which leads to a large free energy.

If the electron has a non-zero mass, the second conservation law is explicitly broken, so the axion rotation should be eventually washed out and all of the charges should be transferred into helical photons. This occurs through the washout of the electron chiral asymmetry with a rate $\sim \alpha m_e^2/T$. Although this rate can be easily much larger than the Hubble expansion rate, since the electron chiral asymmetry is much smaller than the charge in the axion rotation, the washout rate of the axion rotation is suppressed,
\begin{align}
\label{eq:wo_before}
\Gamma_{\rm wo} \simeq \frac{T^2}{f_a^2} \frac{\alpha m_e^2}{T}.
\end{align}

In the SM or the MSSM, there are many fermions with non-zero hypercharges. The washout rate of the axion rotation is dictated by the one with the least chiral symmetry breaking, namely, the electron. After the EWPT, the chiral symmetry breaking is dominantly given by the electron mass $m_e$, so the washout rate is given by Eq.~\eqref{eq:wo_before}.  Before the electroweak phase transition, the chiral symmetry breaking is given by the electron Yukawa coupling $y_e$, so
\begin{align}
\label{eq:wo_after}
\Gamma_{\rm wo} \simeq \frac{T^2}{f_a^2} \alpha_2y_e^2 T,
\end{align}
where $\alpha_2$ is the fine-structure constant of the weak interaction.

In the above discussion, we assume the radial mode is at the minimum, $r\simeq N_{\rm DW} f_a$. To discuss the era where $r \gg N_{\rm DW}f_a$, we may replace $\dot{\theta}$ with $N_{\rm DW} \sqrt{V'(r)/r}$ and $f_a$ with $r$. The evolution of the charge in the axion rotation is parameterized by $r$ rather than $\dot{\theta}$. One can confirm that these rates are much smaller than the Hubble expansion rate throughout the entire evolution, so the axion rotation is never washed out and Eq.~\eqref{eq:happ} gives an accurate estimate of the helicity density obtained when the CPI is fully efficient, even below the equilibrium temperature of the electron Yukawa.

\section{Reynolds numbers}
\label{app:reynolds}

The survival of the hypermagnetic fields until the electroweak phase transition requires the triggering of an inverse cascade in the turbulent regime of magnetohydrodynamics, which shifts the helicity to larger length scales, thus protecting it from diffusion. To check this, we evaluate the kinetic and magnetic Reynolds numbers, with the turbulent regime corresponding to $\text{Re}_\text{mag}, \text{Re}_\text{kin} > 1$. Assuming radiation domination and equipartition between kinetic and magnetic energy (see e.g.~\cite{Durrer:2013pga} and references therein),
\begin{align}
 \text{Re}_\text{mag}  \gtrsim \left[ \frac{\sigma_Y c_\lambda}{H} \left( \frac{\rho_B}{\rho_\text{th}} \right)^{1/2} \right]_{T = T_S}   \sim \frac{10^3}{\sqrt{|c_5|}}\,,
  \label{eq:reynold1}
\end{align}
with $\rho_\text{th}$ denoting the energy density of the thermal plasma and the energy density in the hypermagnetic field is given by $\rho_B \simeq h H / c_\lambda$ for a maximally helical field configuration with a typical correlation length parametrized by $c_\lambda$. 
The square bracket is evaluated when $h$ is dominantly produced at $T = T_S$ since it scales as $1/T$ at $T < T_S$ and thus the Reynolds number only increases after this point.
In the second step, we have used $c_\lambda = H/k_\text{CPI}$ with $k_\text{CPI} = \alpha_Y \mu_{Y,5}/\pi$ and we have inserted the expression for the helicity from Eq.~\eqref{eq:h}.
Note that the final result is independent of $m_\sx$ and $f_a$.

Similarly, for the kinetic Reynolds number,
\begin{align}
 \text{Re}_\text{kin} \gtrsim \left[\frac{c_\lambda}{H \nu}  \left( \frac{\rho_B}{\rho_\text{th}} \right)^{1/2} \right]_{T = T_S} 
  \sim \frac{4}{\sqrt{|c_5|}}
  \label{eq:reynold2}
\end{align}
with $\nu \simeq 10/T$ parametrizing the viscosity. If $\text{Re}_\text{kin} < 1$, we should evaluate the magnetic Reynolds number in the viscous instead of in the equipartition regime. This yields
\begin{align}
 \text{Re}_\text{mag}^\text{visc}  \gtrsim \left[ \frac{\sigma_Y c_\lambda^2}{H^2 \nu} \frac{\rho_B}{\rho_\text{th}} \right]_{T = T_S} 
 \sim \frac{7 \cdot 10^3 }{ |c_5|} \,,
 \label{eq:reynold3}
\end{align}
which is also safely larger than unity. We conclude that if the CPI is triggered, the helical hypermagnetic fields survive until the electroweak phase transition. See Ref.~\cite{Schober:2017cdw} for numerical simulations. See Ref.~\cite{Domcke:2019mnd} for more discussion on these analytical estimates.

In the evaluation of the Reynolds numbers above, we have inserted the maximal helicity obtained when the CPI is completed, as given by Eq.~\eqref{eq:h}. To obtain the correct baryon asymmetry, we require the CPI to be marginally efficient, with the helicity and correspondingly the Reynolds numbers suppressed. In particular, we find the kinetic and magnetic Reynolds number at $T = T_S$ to be smaller than unity, which would indicate a washout of the helicity through diffusion processes before the electroweak phase transition. This indicates that taking into account diffusion, we expect the correct baryon asymmetry is obtained for values of $m_\sx$ slightly above the estimates given in Eq.~\eqref{eq:mS_CPI} and \eqref{eq:mS_CPI_K}, respectively.

\bibliographystyle{utphys}
\bibliography{refs}

\end{document}